\documentclass[%
reprint,
superscriptaddress,
nofootinbib,
amsmath,amssymb,
aps,
prb,
]{revtex4-2}
\usepackage{booktabs}
\usepackage{graphicx}% Include figure files
\usepackage{dcolumn}% Align table columns on decimal point
\usepackage{bm}% bold math
\usepackage{chemformula}
\usepackage{hyperref}% add hypertext capabilities
\usepackage{amsfonts,amssymb}
\usepackage{makecell}
\usepackage{tabularx}% Table width
\usepackage{amsmath}% Higher math
\usepackage{float}
\begin{document}
	\title{Continuous spin excitations in the three-dimensional frustrated magnet \ch{K_2Ni_2(SO_4)_3}}
	\author{Weiliang~Yao}
	\email{weiliangyao@outlook.com}
	\altaffiliation{Present address: Department of Physics and Astronomy, Rice University, Houston, TX 77005, USA}
	\affiliation{Department of Physics and Astronomy, University of Tennessee, Knoxville, TN 37996, USA}
	\author{Qing~Huang}
	\affiliation{Department of Physics and Astronomy, University of Tennessee, Knoxville, TN 37996, USA}
	\author{Tao~Xie}
	\affiliation{Neutron Scattering Division, Oak Ridge National Laboratory, Oak Ridge, TN 37831, USA}
	\author{Andrey~Podlesnyak}
	\affiliation{Neutron Scattering Division, Oak Ridge National Laboratory, Oak Ridge, TN 37831, USA}
	\author{Alexander~Brassington}
	\affiliation{Department of Physics and Astronomy, University of Tennessee, Knoxville, TN 37996, USA}
	\author{Chengkun~Xing}
	\affiliation{Department of Physics and Astronomy, University of Tennessee, Knoxville, TN 37996, USA}
	\author{Ranuri~S.~Dissanayaka~Mudiyanselage}
	\affiliation{Department of Chemistry and Chemical Biology, Rutgers University, Piscataway, NJ 08854, USA}
	\author{Weiwei~Xie}
	\affiliation{Department of Chemistry and Chemical Biology, Rutgers University, Piscataway, NJ 08854, USA}
	\affiliation{Department of Chemistry, Michigan State University, East Lansing, MI 48824, USA}
	\author{Shengzhi~Zhang}
	\affiliation{National High Magnetic Field Laboratory, Los Alamos National Laboratory, Los Alamos, NM 87545, USA}
	\author{Minseong~Lee}
	\affiliation{National High Magnetic Field Laboratory, Los Alamos National Laboratory, Los Alamos, NM 87545, USA}
	\author{Vivien~S.~Zapf}
	\affiliation{National High Magnetic Field Laboratory, Los Alamos National Laboratory, Los Alamos, NM 87545, USA}
	\author{Xiaojian~Bai}
	\affiliation{Department of Physics and Astronomy, Louisiana State University, Baton Rouge, LA 70803, USA}
	\author{D.~Alan~Tennant}
	\affiliation{Department of Physics and Astronomy, University of Tennessee, Knoxville, TN 37996, USA}
	\author{Jian~Liu}
	\affiliation{Department of Physics and Astronomy, University of Tennessee, Knoxville, TN 37996, USA}
	\author{Haidong~Zhou}
	\email{hzhou10@utk.edu}
	\affiliation{Department of Physics and Astronomy, University of Tennessee, Knoxville, TN 37996, USA}
	
	\date{\today}
	
	\begin{abstract}
		Continuous spin excitations are widely recognized as one of the hallmarks of novel spin states in quantum magnets, such as quantum spin liquids (QSLs). Here, we report the observation of such kind of excitations in \ch{K_2Ni_2(SO_4)_3}, which consists of two sets of intersected spin-1 (\ch{Ni^{2+}}) trillium lattices. Our inelastic neutron scattering measurement on single crystals clearly shows a dominant excitation continuum, which exhibits a distinct temperature-dependent behavior from that of spin waves, and is rooted in strong quantum spin fluctuations. Further using the self-consistent-gaussian-approximation method, we determined the fourth- and fifth-nearest neighbor exchange interactions are dominant. These two bonds together form a unique three-dimensional network of corner-sharing tetrahedra, which we name as ``hyper-trillium'' lattice. Our results provide direct evidence for the existence of QSL features in \ch{K_2Ni_2(SO_4)_3} and highlight the potential for the hyper-trillium lattice to host frustrated quantum magnetism.
	\end{abstract}
	
	\maketitle
	
\begin{figure}[b!]
		\centering{\includegraphics[clip,width=8.5cm]{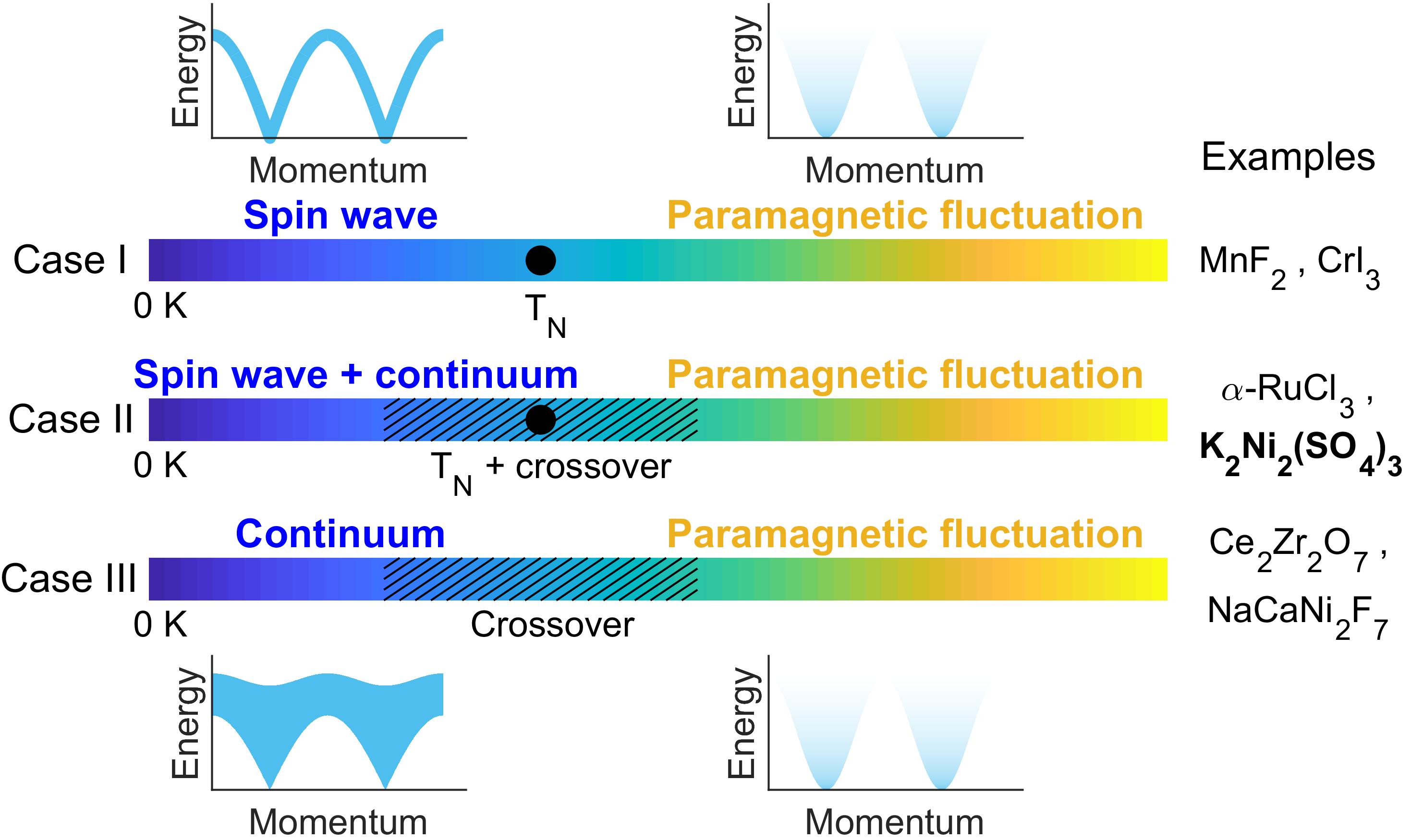}}
		\caption{Schematic of temperature-dependent behaviors for three types of spin excitations observed in magnetic materials with localized moments. Insets display the energy versus momentum relations. The black dots and hatched area denote the ordering temperature ($e.g.$, T$_\mathrm{N}$) and crossover region, respectively. The materials representative of each case are listed on the right.}
		\label{fig1}
	\end{figure}

	For conventional insulating magnets, spins usually order at a finite temperature ($e.g.$, T$\rm_N$), below which sharp spin waves emerge due to the propagation of spin fluctuations \cite{Bloch1930} (case I of Figure \ref{fig1}). When warming above the ordering temperature, spin waves disappear with paramagnetic fluctuations remaining. In contrast, quantum spin liquids (QSLs) have other spectroscopic features due to long-range quantum entanglement \cite{BalentsNature2010,WenNPJQM2019,WulferdingJPCM2019,KnolleARCMP2019}. Namely, the spins can fractionalize into fermionic quasi-particles so that can only be detected in pairs by spectroscopic methods, which exhibit a continuous excitation spectrum \cite{BalentsNature2010,WenNPJQM2019,WulferdingJPCM2019,KnolleARCMP2019} (case III of Fig. \ref{fig1}). For example, \ch{Ce_2Zr_2O_7} \cite{GaoNP2019,GaudetPRL2019,SmithPRX2022,GaoPRB2022} and \ch{NaCaNi_2F_7} \cite{KrizanPRB2015,PlumbNP2019} are two representative materials showing remarkable continuous spin excitations that are related to QSL. Due to the lacking of an ordering transition, the continuum transfers to paramagnetic spectrum through a crossover when warming up \cite{SavaryRPP2017} (Fig. \ref{fig1}).
	
	However, when geometric spin frustration and/or competing interactions are significant, some spin-ordered magnets will still exhibit remarkable QSL signatures including excitation continuum (case II of Fig. \ref{fig1}). The interplay of multiple ingredients suggests their magnetic properties may be susceptible to external tuning parameters, such as chemical doping \cite{ZungerCM2021}, magnetic field \cite{MaCPB2018}, and pressure \cite{BiesnerCryst2020}, which sets them apart from the above two categories. The Kitaev spin liquid candidate $\alpha$-\ch{RuCl_3} has been studied as a celebrated example in the case II, which hosts both spin waves and continuous spin excitations in its antiferromagnetic ordered state \cite{BanerjeeNM2016,BanerjeeScience2017,DoNP2017,BanerjeeNPJQM2018,HanArxiv2022}. On the one hand, a rod-like magnetic continuum at Brillouin zone center was observed by neutron, Raman, and terahertz spectroscopies \cite{BanerjeeNM2016,BanerjeeScience2017,DoNP2017,BanerjeeNPJQM2018,HanArxiv2022,SandilandsPRL2015,WulferdingNC2020,LittlePRL2017}, which has been widely viewed as a ``smoking-gun'' for fractionalized Majorana fermions. On the other hand, the long-range magnetic order and spin waves present at zero field can be fully suppressed by an in-plane magnetic field, resulting a QSL state before partially magnetic polarization \cite{SearsPRB2017,BaekPRL2017,BanerjeeNPJQM2018,Balz2019}.
	
	\begin{figure}[t!]
		\centering{\includegraphics[clip,width=8.5cm]{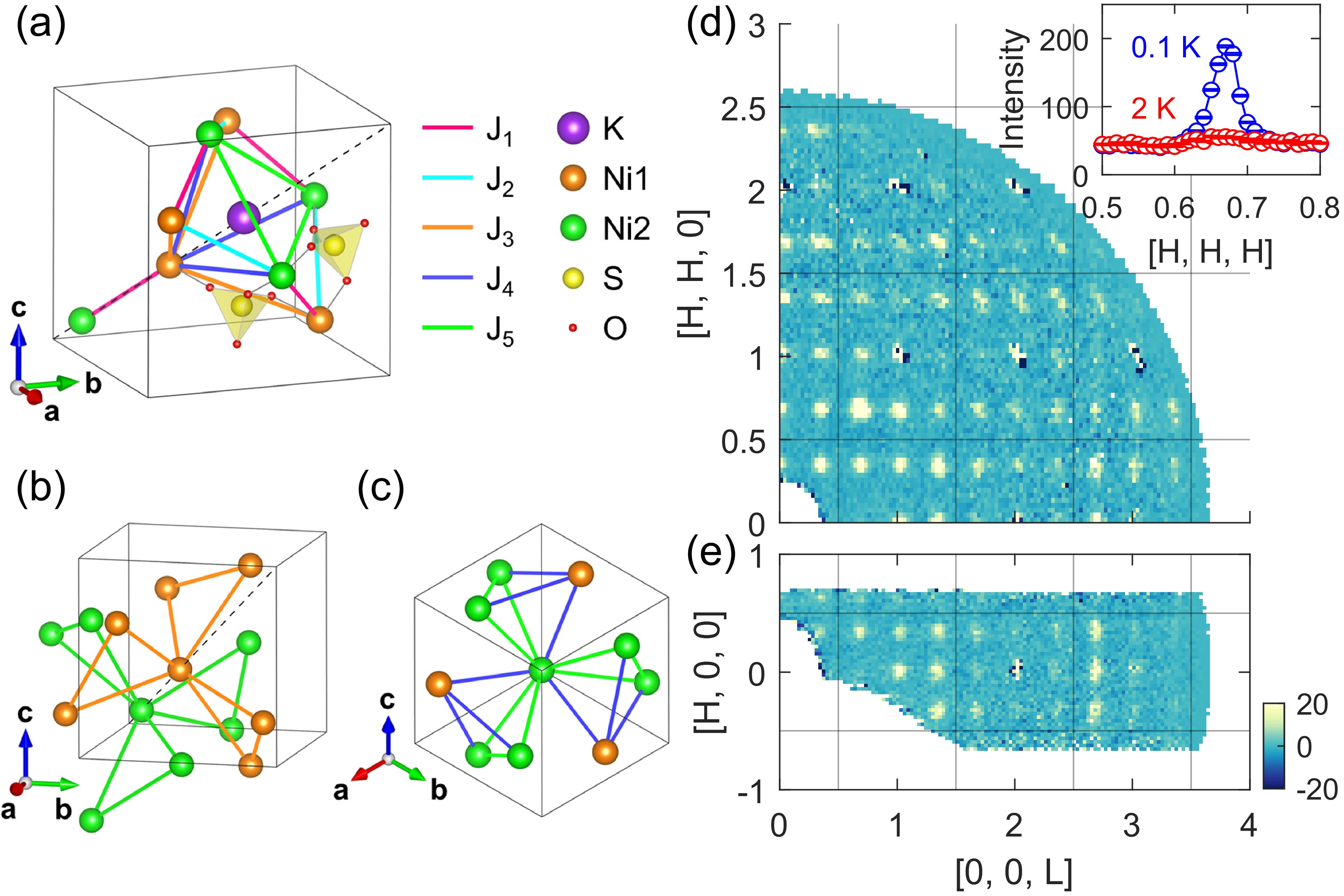}}
		\caption{(a) Crystal structure of \ch{K_2Ni_2(SO_4)_3}. For clarity, only two \ch{SO_4^{2-}} groups (depicted as yellow tetrahedra) and one \ch{K^{+}} ion are shown explicitly. Colored lines indicate exchange interactions of up to $J_5$ between \ch{Ni^{2+}} ions. The dashed line represents the body diagonal along the [1, 1, 1] direction of the cubic unit cell. (b) Trillium lattices of Ni1 and Ni2 formed by $J_3$ and $J_4$ bonds, respectively. (c) Hyper-trillium lattice (see text) formed by $J_4$ and $J_5$ bonds. These structural illustrations were generated using VESTA \cite{MommaJAC2011}. (d) and (e) Magnetic diffraction patterns of the ($H$, $H$, $L$) and ($H$, 0, $L$) planes at 0.1 K, with data at 2 K subtracted as background. The solid lines represent Brillouin zone boundaries. The inset of (d) shows the intensity around (2/3, 2/3, 2/3) at 0.1 K and 2 K.}
		\label{fig2}
	\end{figure}
	
	More recently, the langbeinite compound \ch{K_2Ni_2(SO_4)_3} has been proposed to be another field-induced QSL based on thermodynamic and spectroscopic measurements \cite{ZivkovicPRL2021}. With a cubic structure, \ch{K_2Ni_2(SO_4)_3} has two sets of spin-1 \ch{Ni^{2+}}-trillium lattice interconnecting in three-dimensional (3D) space [Figure \ref{fig2}(a) and (b)] \cite{ZivkovicPRL2021,SpeerPCM1986}. Although magnetic phase transitions to spin-ordered states have been identified in this compound, prominent quantum spin fluctuations are evidenced by a broad hump in magnetic specific heat and a plateau of relaxation rate in muon spin spectroscopy \cite{ZivkovicPRL2021}. Appreciable quasielastic scattering and continuum-like excitations were also respectively observed by neutron diffraction and inelastic neutron scattering (INS) on powder sample \cite{ZivkovicPRL2021}. Moreover, a moderate magnetic field $\sim$4 T can fully suppress the ordered spin components and drive the system into a QSL state \cite{ZivkovicPRL2021}. These findings suggest that the magnetic order in \ch{K_2Ni_2(SO_4)_3} is fragile and it may approximate to a QSL at zero field. However, the nature of its spin dynamics is less clear due to the limitation of the powder data. In particular, whether the observed spin excitations are intrinsically continuous or simply powder-averaged spin waves is the major unknown.
	
	In this work, we present an INS study on large and high-quality \ch{K_2Ni_2(SO_4)_3} single crystals. We find that although a long-range magnetic order develops below T$\rm_N\approx$ 1.1 K, its spin excitations are continuous even at temperatures well below T$\rm_N$, such as down to 0.1 K. By studying these excitations over a temperature range covering almost three orders of magnitude, we conclude that they are distinct from spin-wave excitations observed in conventional magnets but similar to those continuous spin excitations in studied QSL candidates. With the self-consistent-gaussian-approximation (SCGA) method, we determined the fourth- and fifth-nearest-neighbor exchange interactions are dominant, which in together construct a hitherto uncovered structure - the ``hyper-trillium'' lattice. Our study on \ch{K_2Ni_2(SO_4)_3} therefore shows another rare example for the existence of QSL features amid a spin-ordered state.
	
	Single crystals of \ch{K_2Ni_2(SO_4)_3} were prepared with a self-flux method \cite{ZivkovicPRL2021,SM}. In our INS experiment, 9 pieces of single crystals with a total mass of $\sim$6 grams were cut and coaligned with the ($H$, $H$, $L$) plane being put in horizontal. The experiment was performed in the Cold Neutron Chopper Spectrometer (CNCS) installed at Spallation Neutron Source, Oak Ridge National Laboratory \cite{EhlersRSI2011}. Throughout the experiment, an incident neutron energy of 3.32 meV was employed in the high flux mode. A dilution refrigerator insert was used to provide a base temperature of 0.1 K. We rotated the sample along the vertical [1, -1, 0] direction about 150$^\circ$ to fully cover one quadrant of the ($H$, $H$, $L$) plane. To present intensity maps of this plane, we symmetrized the data according to the crystal symmetry of \ch{K_2Ni_2(SO_4)_3}. Data at six temperatures (0.1 K, 0.9 K, 2 K, 10 K, 20 K, and 80 K) were collected, which were reduced and analyzed with Horace \cite{EwingsHorace2016}. The neutron scattering intensity was converted to absolute unit based on structural Bragg peaks as described in \cite{SM}.
	
	Fig. \ref{fig2}(d) and (e) show elastic magnetic scattering maps of the ($H$, $H$, $L$) and ($H$, 0, $L$) planes at 0.1 K. We can identify magnetic Bragg peaks at positions that are indexed by (1/3, 0, 0), (1/3, 1/3, 0), and (1/3, 1/3, 1/3), which is consistent with the previous report based on the powder sample \cite{ZivkovicPRL2021}. Additionally, we also find intensity at Brillouin zone centers [$e.g.$, at (1, 0, 0)], which might be caused by magnetic multiple scattering, or an extra magnetic wave vector $\textbf{k}$ = (0, 0, 0). Despite the detailed magnetic structure is beyond this study, the coexistence of multiple propagation vectors indicates a magnetic ground state with several competing phases. We notice that two thermal phase transitions at 0.74 K and 1.14 K were reported previously \cite{ZivkovicPRL2021}. However, we here only observed one transition at $\sim$1.1 K by magnetic susceptibility, and we did not find significant change of the magnetic Bragg peaks around 0.7 K (see \cite{SM} for details).
	
\begin{figure}[t!]
		\centering{\includegraphics[clip,width=8.7cm]{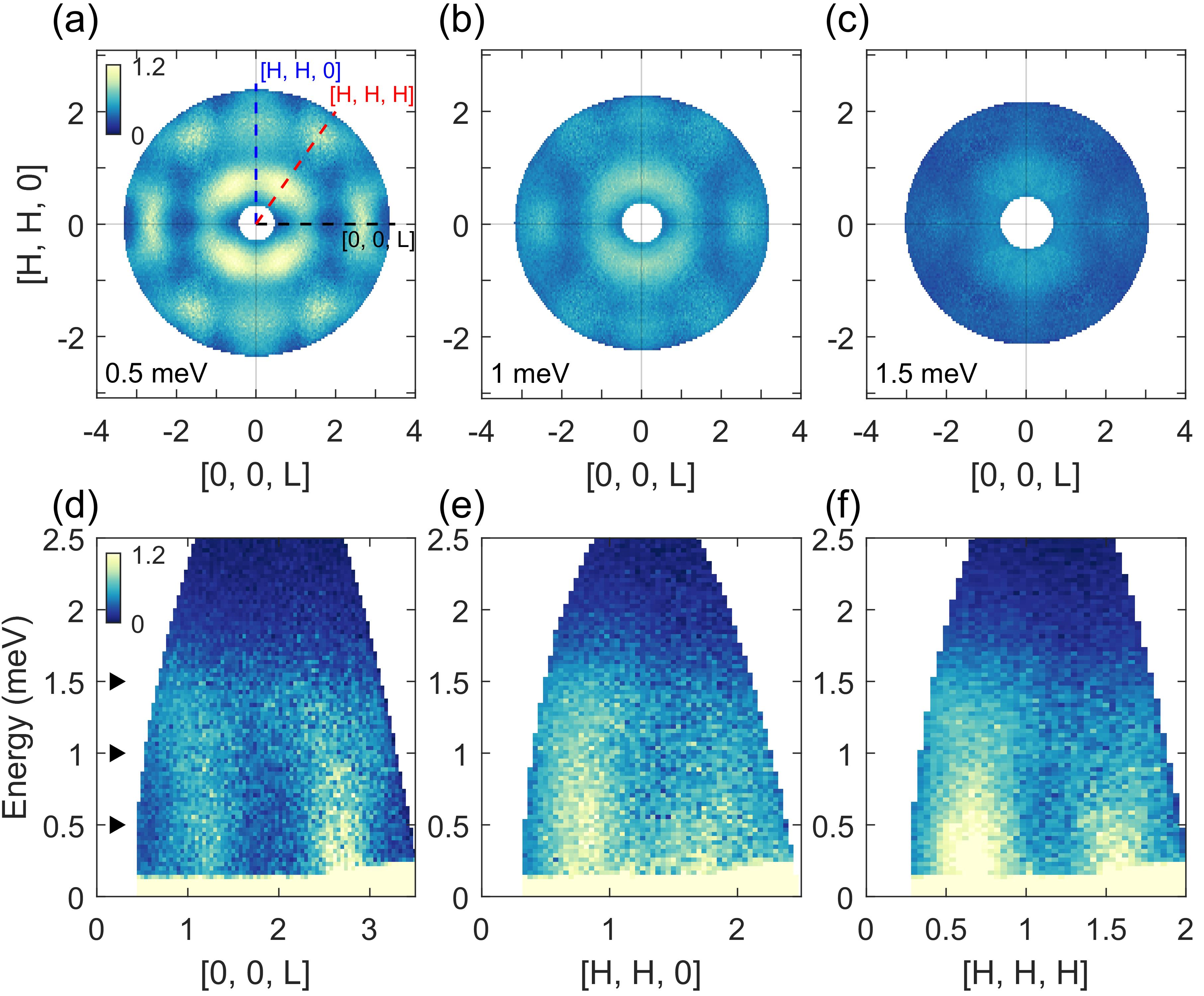}}
		\caption{(a)-(c) Constant energy slices of the ($H$, $H$, $L$) plane at 0.1 K. (d)-(f) Energy dependence of the magnetic continuum along high-symmetric directions [dashed lines in (a)]. The right triangles indicate in (d) energy positions where the slices in (a)-(c) were taken.}
		\label{fig3}
	\end{figure}

	Spin excitations at 0.1 K are presented in Figure \ref{fig3}. Albeit the temperature is only about 0.1T$\rm_N$, we find the excitation spectrum is dominated by a broad continuum. The constant energy slices in Fig. \ref{fig3} (a)-(c) show that the dynamic structure factor reaches its maximum around (2/3, 2/3, 2/3) in the ($H$, $H$, $L$) plane, which corresponds to the strongest magnetic Bragg peak [Fig. \ref{fig2}(d)]. This indicates the continuous spin excitations are intimately related to the underlying magnetic order. According to energy–momentum slices [Fig. \ref{fig3} (d)-(f)], these excitations are gapless and extend up to $\sim$2 meV, which is consistent with the Weiss temperature [$\Theta_{\mathrm{CW}}$ = -29.6(1) K] \cite{SM} and the reported powder INS data \cite{ZivkovicPRL2021}.
	
\begin{figure}[t!]
		\centering{\includegraphics[clip,width=8.7cm]{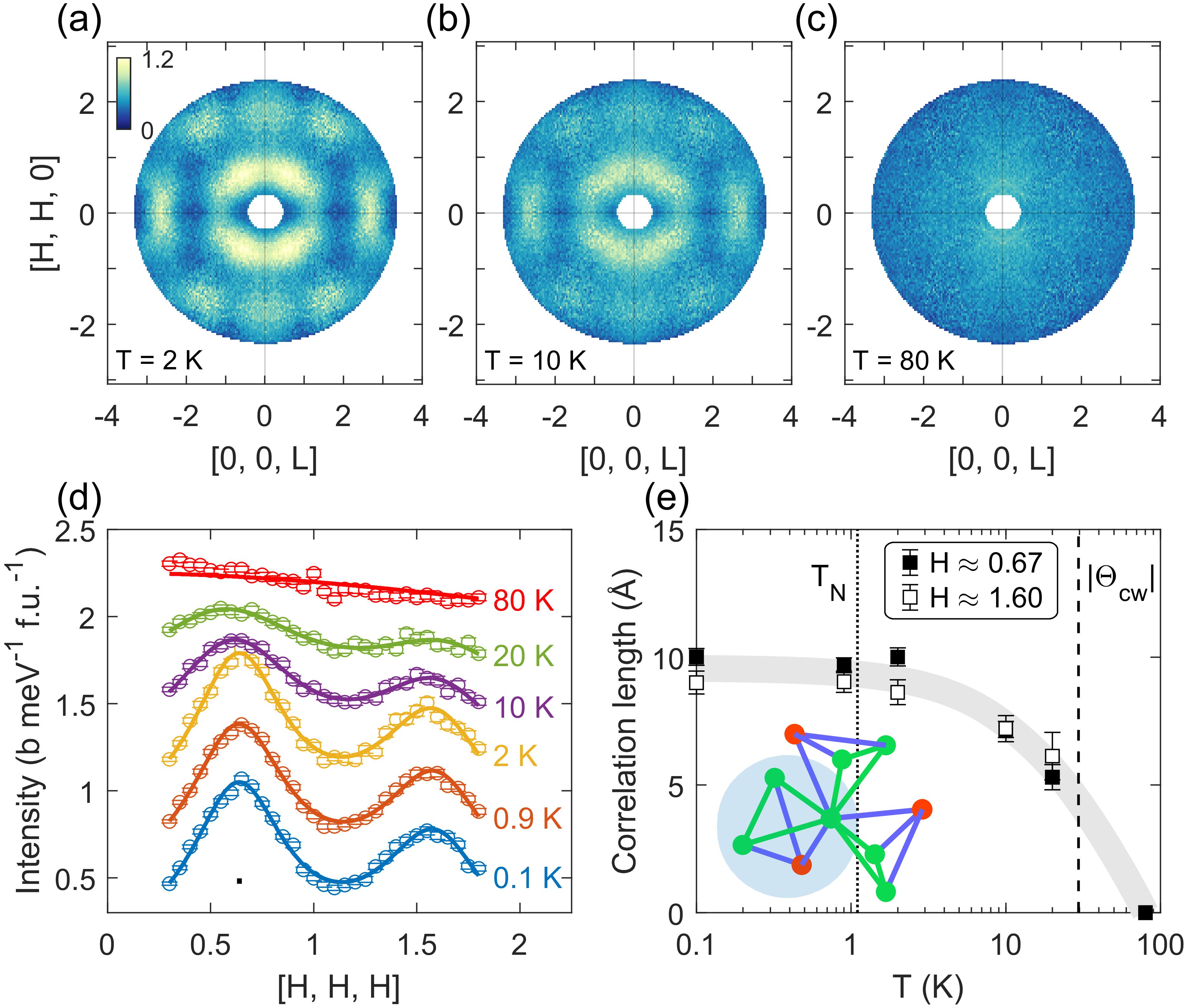}}
		\caption{(a)-(c) Constant energy slices for 0.5 meV at 2 K, 10 K, and 80 K. (d) Momentum dependence of the intensity along [$H$, $H$, $H$] obtained by integrating over $E$ = [0.2, 1.0] meV. Data above 0.1 K are vertically offset for clarity. Solid curves are fits to the data as described in the text. Short horizontal bar indicates the momentum resolution. (e) Temperature dependence of the spin correlation length determined from the peak widths in (d). Bold grey line is a guide to the eyes. The dotted and dashed vertical lines mark the T$\rm_N$ and $|\Theta_{\mathrm{CW}}|$, respectively. The inset shows the correlation sphere (shown in light blue) at 0.1 K with respect to tetrahedral units.}
		\label{fig4}
	\end{figure}

	Upon warming to 2 K, the scattering pattern is largely intact [Figure \ref{fig4}(a)], despite the fact that the long-range magnetic order has faded away [see the inset of Fig. \ref{fig2}(d)]. Similar scattering pattern is still apparent at 10 K [Fig. \ref{fig4}(b)], and finally becomes featureless at 80 K [Fig. \ref{fig4}(c)], where the intensity decays with the momentum transfer by following the magnetic form factor of \ch{Ni^{2+}} [\cite{SM} and Fig. \ref{fig4}(d)]. Energy–momentum slices of other five temperatures are presented in Fig. S6 of \cite{SM}. This temperature dependence behavior further confirms that the observed signal is from magnetic scattering. By analyzing the data at 0.1 K and 2 K, we find the spectral weight in the elastic channel is less than 10\% of the total \cite{SM}. On the other hand, for a conventional spin-1 Heisenberg magnet, half of the spectral weight is expected to be elastic. This feature is an indicative of strong quantum spin fluctuations in \ch{K_2Ni_2(SO_4)_3} \cite{PlumbNP2019}. Similar observation has been made on the QSL candidate \ch{NaCaNi_2F_7} \cite{PlumbNP2019}, in which $\sim$90\% of the neutron scattering spectral weight forms continuous spin excitations. 
	
	In Fig. \ref{fig4}(d), we show constant energy cuts (of $E$ = [0.2, 1.0] meV) along [$H$, $H$, $H$]. At temperatures below 80 K, there are two broad peaks at $H \approx$ 0.67 and $H \approx$ 1.60, which can be well fitted with a double-Lorentzian profile multiplied with the square of magnetic form factor (solid curves). Based on the fitted peak widths, we extracted the spin correlation lengths at various temperatures \cite{MacDougallPNAS2011,YoungPRB2013}, which are presented in Fig. \ref{fig4}(e). Since there is no peak can be resolved for the data at 80 K, we fitted them only with the square of magnetic form factor, and set the correlation length to be zero. In spite of the fact that the two peaks are rather different in intensity, the correlation lengths deduced from them are basically the same, which therefore can be regarded as a representative parameter. 
	
	The spin correlation length at 0.1 K ($\xi_0$) is about 9.5 \AA. It is much smaller than the one estimated based on the magnetic Bragg peak [inset of Fig. \ref{fig1}(d)], which is about 197 \AA. The greatly reduced correlation length for the inelastic signal indicates its short-range nature. Interestingly, the correlation sphere defined by $\xi_0$ approximately covers the tetrahedral unit formed by Ni1 and Ni2 [inset of Fig. \ref{fig4}(e)], as will be discussed below. There are two noteworthy features in the temperature dependence of this correlation length. First, it is only below $\sim$$|\Theta_{\mathrm{CW}}|$ that the spin correlation significantly establishes, which reflects the fact that the exchange interactions govern the spin dynamics. Second and more importantly, it is basically unchanged when the temperature goes across T$\rm_N$. This behavior is different from conventional spin waves, where the correlation length is expected to decrease on approaching the ordering temperature \cite{LynnPRB1975,LynnPRB1981}. Such insensitivity to the T$\rm_N$ reveals that the major spin dynamics of \ch{K_2Ni_2(SO_4)_3} is distinct from spin waves. Instead, the thermal evolution of the correlation length resembles those studied QSLs \cite{XuScience2007,ClarkNP2019}, indicating the observed continuous spin excitations may origin from a QSL state.
	
\begin{figure}[t!]
		\centering{\includegraphics[clip,width=8.0cm]{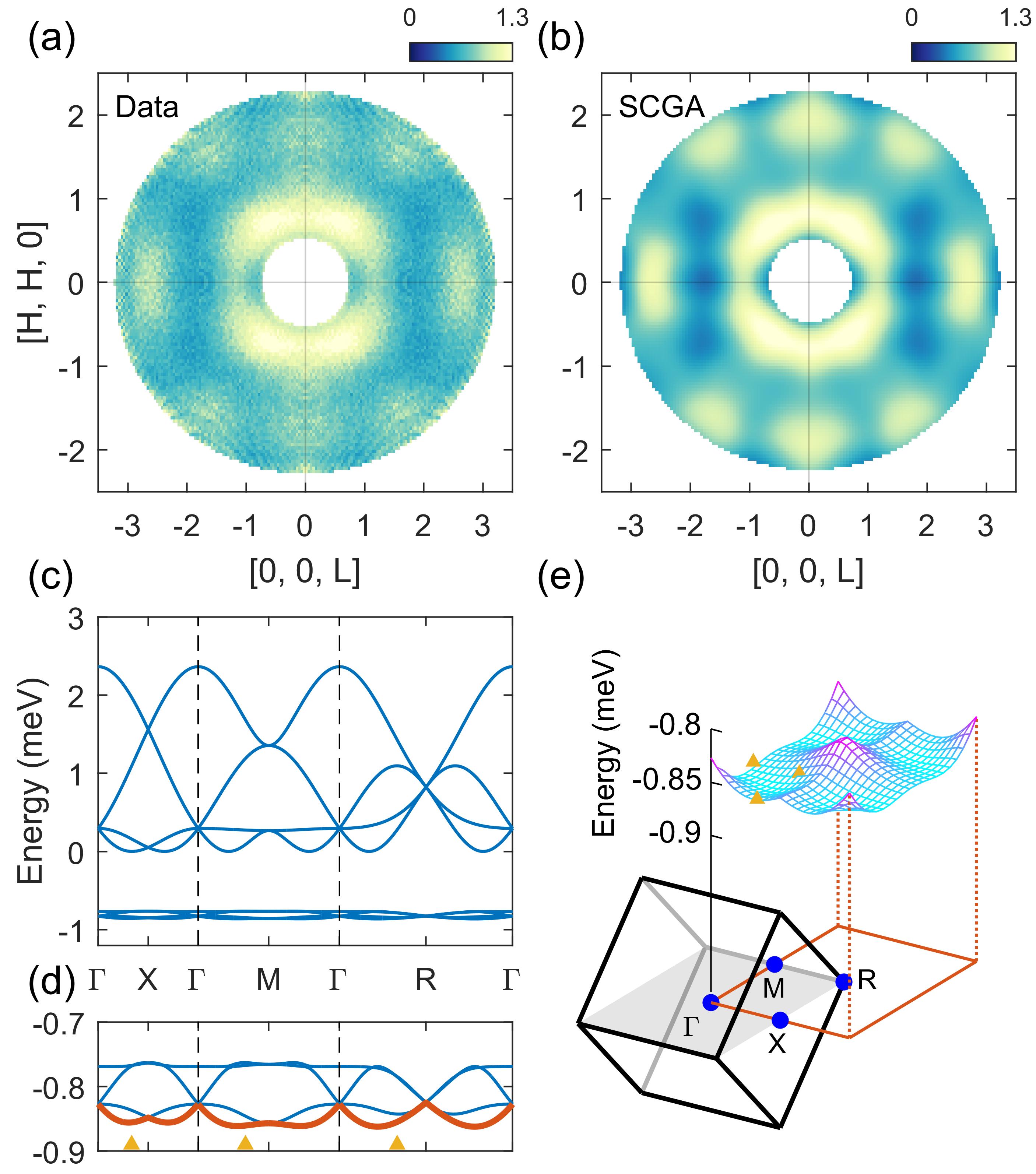}}
		\caption{(a) Energy-integrated intensity map (from 0.15 meV to 2 meV) of the ($H$, $H$, $L$) plane at 2 K. (b) The corresponding scattering pattern calculated using the SCGA method. (c) Dispersion curves for the interaction matrix along high-symmetric directions in the Brillouin zone. (d) A zoom-in view of four energy bands around -0.8 meV in (c). The energy band with the lowest energy is highlighted in orange. (e) The lowest energy band shown in the ($H$, $H$, $L$) plane. The cube is the first Brillouin zone, with high-symmetric points indicated by blue dots. The upper triangles in (d) and (e) indicate positions of the three magnetic wave vectors observed experimentally.}
		\label{fig5}
	\end{figure}

	Next, we establish the major exchange interactions of \ch{K_2Ni_2(SO_4)_3} with the SCGA method, which has been widely used to determine magnetic exchange interactions in frustrated magnets \cite{ConlonPRB2010,BentonJPSJ2015,PlumbNP2019,BaiPRL2019,PaddisonPRL2020,GuPRB2022,GaoPRL2022,GaoPRB2022}. In our calculation, we used the following isotropic Heisenberg Hamiltonian as a starting point
	\begin{equation}
		H = \frac{1}{2}\sum_{n=1}^5J_n\sum_{i,j}\textbf{S}_i \cdot \textbf{S}_j,
	\end{equation}
	where $J_1$, $\cdots$, $J_5$ are exchange interactions up to fifth-nearest-neighbor. Figure \ref{fig5}(a) shows the energy integrated intensity map at 2 K, which is approximately proportional to the magnetic structure factor. By fitting this spectrum with the SCGA method, we determined the exchange interactions as: $J_1$ = -0.03(2) meV; $J_2$ = 0.00(1) meV; $J_3$ = 0.01(1) meV; $J_4$ = 0.47(2) meV; $J_5$ = 0.26(2) meV. The details of the SCGA calculation can be found in \cite{SM}. With these parameters, the calculated intensity map is presented in Fig. \ref{fig5}(b), which reproduces most of the features in our data. We point out that a single-ion anisotropic term is usually allowed for S = 1 \ch{Ni^{2+}} ions, yet in our case, the inclusion of it cannot significantly improve the fitting \cite{SM}. Therefore, this Hamiltonian can be regarded as a minimum effective model for the magnetism of \ch{K_2Ni_2(SO_4)_3}.
	
	To inspect the magnetic ground state, we further calculated the energy bands of the interaction matrix \cite{SM}, where momentum positions with minimum energy predict the magnetic ordering wave vector in the mean field level \cite{ReimersPRB1991}. As presented in Fig. \ref{fig5}(c), there are four low energy bands that are extremely flat, which naturally accounts for the highly frustrated nature of this system. Looking closer on those flat bands, we find the observed ordering wave vectors locate around the valley of the lowest-energy band [Fig. \ref{fig5}(d) and (e)]. It again suggests that the determined Hamiltonian is a good approximation, and the system features a variety of competing states with very close energy.
	
	For this set of parameters, it is noteworthy that $J_4$ and $J_5$ are significant, and other exchange interactions are negligibly small. Structurally, the bonds of $J_4$ and $J_5$ together form a 3D lattice with corner-sharing tetrahedra [Fig. \ref{fig2}(c)], reminiscent of the pyrochlore lattice \cite{GardnerRMP2010}. Due to the slight difference between the bond lengths of $J_4$ and $J_5$ (by $\sim$0.1 \% \cite{SM}), Ni1 and Ni2 are fundamentally inequivalent in this 3D network. Specifically, three tetrahedra share one corner at the Ni2 site, while each Ni1 only belongs to one tetrahedron, which connects to Ni2 through $J_4$. To the best of our knowledge, such kind of lattice has never been reported before. In order to facilitate future studies on this newly identified structure, we here dub it as \textit{hyper-trillium} lattice. Additional structural illustrations showing more tetrahedral units can be found in Fig. S2 of \cite{SM}. It is easy to see that the hyper-trillium lattice inherits the 3D connection from the trillium lattice [Fig. \ref{fig2}(b)]. The common magnetic propagation wave vector (1/3, 0, 0) for both the trillium lattice and \ch{K_2Ni_2(SO_4)_3} further hints their close relationship \cite{HopkinsonPRB2006,HopkinsonPRB2007,IsakovPRB2008,BulledPRL2022}. Nevertheless, our finding indicates the hyper-trillium lattice would be a more straightforward model to describe the magnetism of \ch{K_2Ni_2(SO_4)_3}, rather than two sets of trillium lattices.
	
	The prominence of $J_4$ and $J_5$ can be understood from the structural perspective. We first note that all Ni-Ni exchange interactions up to $J_5$ are mediated by the \ch{SO_4^{2-}} group, which form a Ni-O-S-O-Ni super-superexchange path [Fig. \ref{fig2}(a)]. Farther interactions ($J_6$ $\dots$) involve multiple intermediate groups so as to be considerably smaller. For $J_3$, $J_4$, and $J_5$, their exchange paths (Ni-O-S-O-Ni) are more straight than those of $J_1$ and $J_2$, which can be more directly seen from the bond angle of Ni-S-Ni (see Table S3 in \cite{SM}). Moreover, the K$^{+}$ ion locates very closely to the center of the octahedral unit that formed by $J_4$ and $J_5$ [Fig. \ref{fig2}(a)]. This cation may attract more electrons to hop around this region, so as to enhance the exchange interaction. Therefore, the formation of the hyper-trillium lattice can be attributed to the combination of relatively straight exchange paths and cation attraction effect.
	
	In summary, we have observed continuous spin excitations in \ch{K_2Ni_2(SO_4)_3} with INS, which persist well below its T$_\mathrm{N}$. The temperature dependence of these excitations shows a distinct behavior from conventional spin waves, suggesting a close connection to QSL. Using the SCGA method, we have determined that the fourth- and fifth-nearest-neighbor exchange interactions are dominant, which effectively constitute a hyper-trillium lattice that is responsible for the much enhanced spin frustration. Our study not only uncovers decisive QSL features in \ch{K_2Ni_2(SO_4)_3}, but also demonstrates that the hyper-trillium is a new platform to explore frustrated quantum magnetism. As an immediate consequence, it may be applicable to the large number of other langbeinite compounds with rich chemical variants \cite{Gattow1958,McmurdieNIST1971,SpeerPCM1986}, which is expected to stimulate wide-ranging research of interest, as those pyrochlore oxides have done in the past few decades \cite{GardnerRMP2010}. Ultimately, future studies aimed at clarifying the essence of the observed excitation continuum, $i.e.$, whether it is truly related to fractionalized fermionic excitations, are desirable.
	
	\vspace{\baselineskip}             
	\textbf{Acknowledgments} We wish to thank Tianran Chen, Tong Chen, Seung-Hwan Do, Chunruo Duan, Dongliang Gong, Martin Mourigal, and Feng Ye for discussion. This research was supported by the U.S. Department of Energy under grant No. DE-SC0020254. This research used resources at the Spallation Neutron Source, a DOE Office of Science User Facility operated by the Oak Ridge National Laboratory. Part of the work was done in the National High magnetic Field Laboratory, supported by the U.S. Department of Energy, Office of Science, National Quantum Information Sciences Research Centers, Quantum Science Center. The facilities of the National High Magnetic Field Laboratory are supported by the National Science Foundation Cooperative Agreement No. DMR-1644779, and the State of Florida and the U.S. Department of Energy. S. Z. also acknowledges LDRD program at Los Alamos National Laboratory.
	
	%
	%\textbf{Author contributions} W. Y. grew and coaligned the single crystals with the help of Q. H., C. X., and H. Z. W. Y. conducted the neutron scattering experiment with the help of T. X. and A. P., which was designed by W. Y., H. Z., and D. A. T. W. Y. analyzed the neutron scattering data. R. D. D. M. and W. X. performed single crystal XRD measurements and refined the crystal structure. W. Y. and A. B. measured magnetic susceptibility with in-house MPMS. S. Z., M. L., and V. S. Z. performed the high field magnetization measurements. W. Y. performed the calculation and fitting on the consult of X. B. W. Y. and H. Z. prepared the manuscript with the inputs from all authors.
	
	\bibliographystyle{apsrev4-2}
	\bibliography{knso_ins_reference}
	
%%%%%%%%%% Merge with supplemental materials %%%%%%%%%%
\pagebreak
\pagebreak

\widetext
\begin{center}
	\textbf{\large Supplemental Material for ``Continuous spin excitations in the three-dimensional frustrated magnet \ch{K_2Ni_2(SO_4)_3}''}
\end{center}

%%%%%%%%%% Merge with supplemental materials %%%%%%%%%%
%%%%%%%%%% Prefix a "S" to all equations, figures, tables and reset the counter %%%%%%%%%%
\setcounter{equation}{0}
\setcounter{table}{0}
\setcounter{page}{1}
\makeatletter
\renewcommand{\theequation}{S\arabic{equation}}
\renewcommand{\thefigure}{S\arabic{figure}}
%\renewcommand{\bibnumfmt}[1]{[S#1]}
%\renewcommand{\citenumfont}[1]{S#1}
%%%%%%%%%% Prefix a "S" to all equations, figures, tables and reset the counter %%%%%%%%%%

\section{\ch{K_2Ni_2(SO_4)_3} single crystals and characterizations}

\subsection{Single crystal growth method}
\ch{K_2Ni_2(SO_4)_3} single crystals were prepared with a self-flux method. We first prepared \ch{K_2Ni_2(SO_4)_3} polycrystalline sample with solid state reaction method. Starting materials of \ch{K_2SO_4} and \ch{NiSO_4}$\cdot$6\ch{H_2O} were weighed in a molar ratio of 1:2. The mixture was ground and pressed to rod, which was put into a box furnace and heated to 450 $^{\circ}$C in air for 40 hours. The resultant rod was ground and pressed to rod again, following another 40-hour sintering at 450 $^{\circ}$C in air. To grow single crystals, the polycrystalline sample was sealed into an evacuated quartz tube and heated to 850 $^{\circ}$C in a box furnace. The quartz tube was kept in 850 $^{\circ}$C for 20 hours before cooling down to 750 $^{\circ}$C in a rate of 0.5 $^{\circ}$C per hour. Then, the furnace was turned off and the quartz tube was cooled to room temperature. Yellow single crystals in irregular shape can be harvested in the solidified product.

\begin{figure}[h!]
	\centering{\includegraphics[clip,width=11cm]{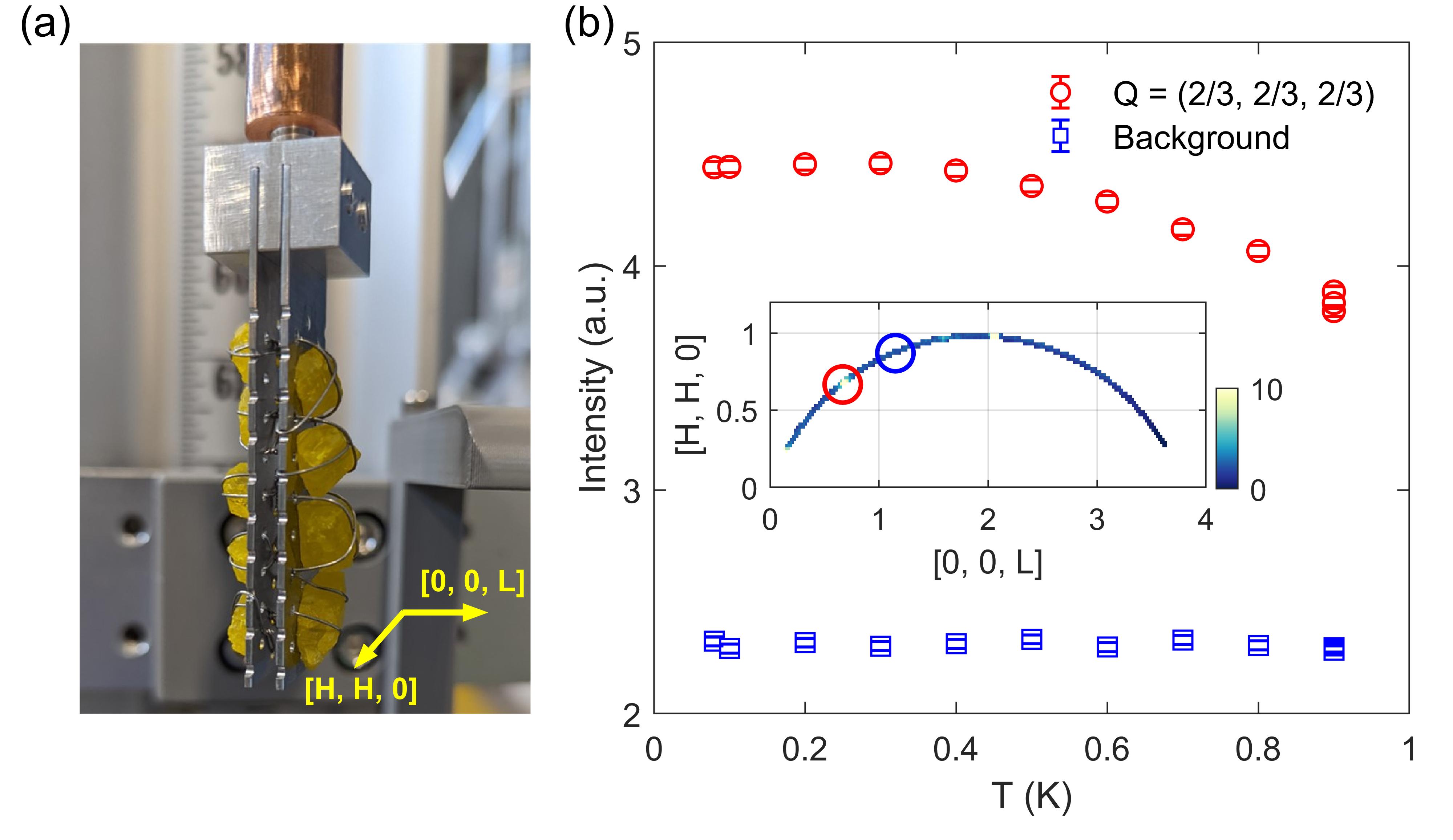}}
	\caption{(a) Single crystal array used in the neutron scattering experiment. (b) Temperature dependence of the intensity at (2/3, 2/3, 2/3) and (0.85, 0.85, 1.15), which are positions of magnetic Bragg peak and background, respectively. Inset shows the covered reciprocal trajectory in this temperature dependence measurement.}
	\label{figs1}
\end{figure}

In our neutron scattering experiment, 9 pieces of large single crystals were cut and coaligned with a Huber Laue diffractometer. The ($H$, $H$, $L$) plane was put in horizontal as the scattering plane [see Fig. \ref{figs1}(a)]. Rocking scans performed on Bragg peaks (1, 1, -1) and (1, 1, 0) indicate a sample mosaic spread of about 4$^{\circ}$ and 2$^{\circ}$, respectively.

\subsection{Single crystal X-ray diffraction and crystal structure}
\begin{figure}[h!]
	\centering{\includegraphics[clip,width=11cm]{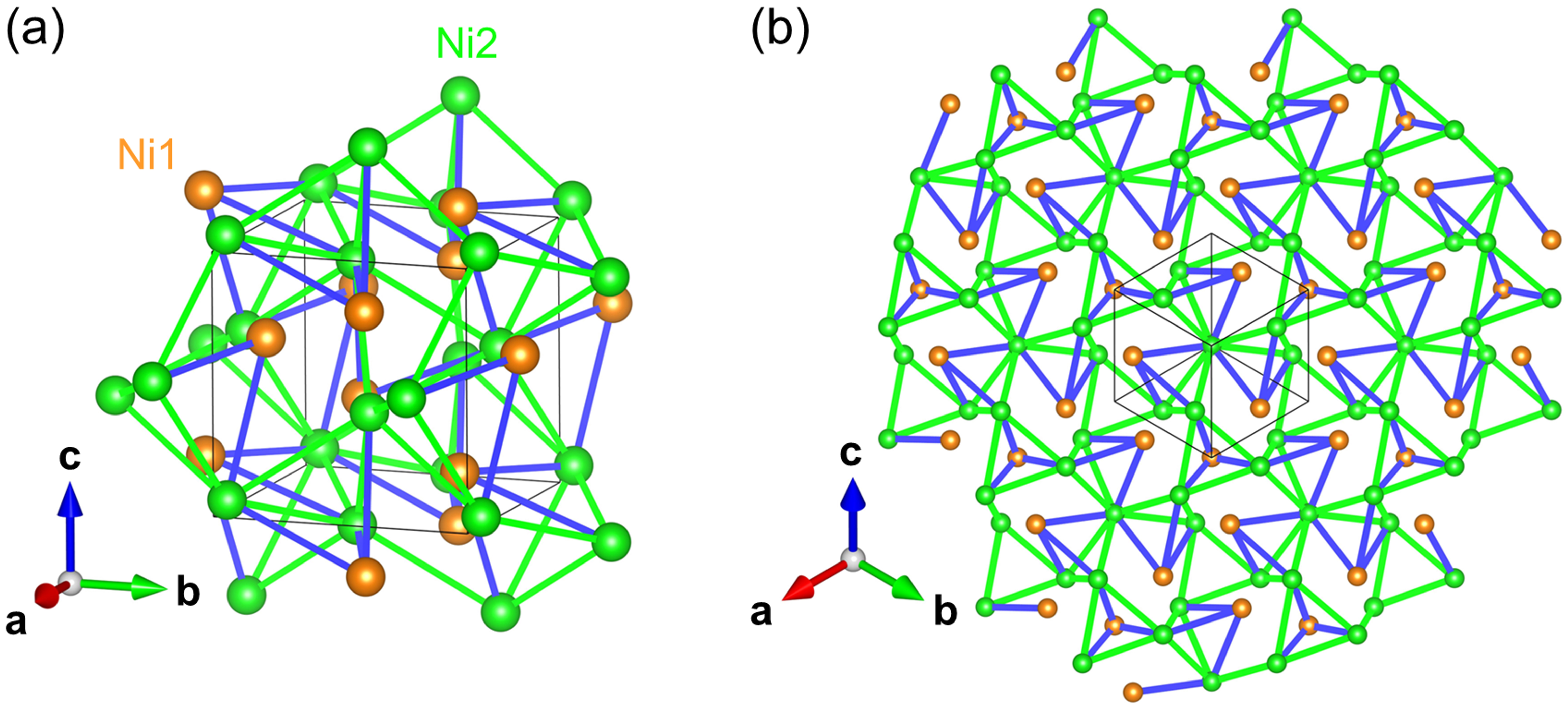}}
	\caption{(a) Hyper-trillium lattice plotted in a range around the cubic unit cell. (b) Hyper-trillium lattice viewed along [1, 1, 1] direction, which is an extended plot of Fig. 2 (c). For simplicity, only one ``layer'' of tetrahedral units is showed.}
	\label{figs2}
\end{figure}

To determine the crystal structure and confirm the chirality of the phase, the single crystal X-ray diffraction measurement was applied for the yellow crystal \ch{K_2Ni_2(SO_4)_3} at room temperature with Mo radiation in a Bruker Eco Quest. The obtained crystal symmetry is cubic with the chiral space group, $P$2$_1$3. The atomic occupancy was refined using the full-matrix least-squares on F$^2$ method. The result shows that no mixture or vacancies were detected in the system. The crystallographic information can be found in Table \ref{tb1} and Table \ref{tb2}. The corresponding bond distance and angle information is listed in Table \ref{tb3}. We used this crystal crystallographic information throughout the paper. In Fig. \ref{figs2}, we show the hyper-trillium lattice structure in more extended ranges.

\subsection{Magnetic susceptibility}
\begin{figure}[b!]
	\centering{\includegraphics[clip,width=16cm]{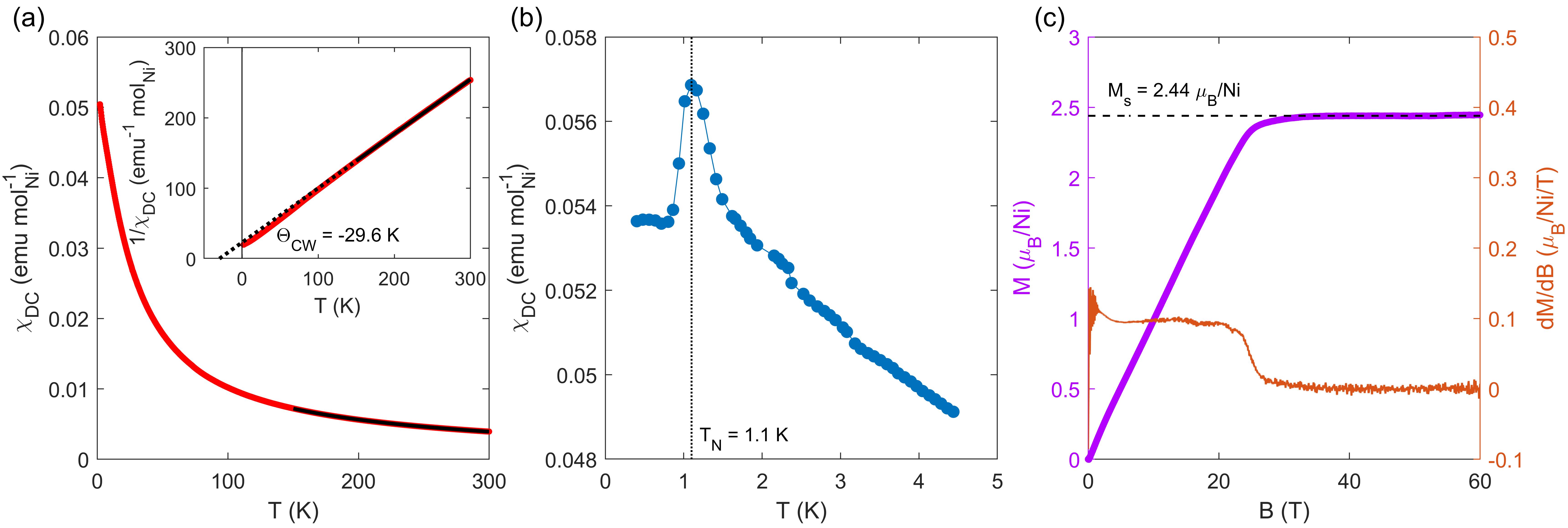}}
	\caption{(a) DC magnetic susceptibility from 2 K to 300 K measured under a magnetic field of 0.2 T applied along [1, 1, 0] direction. The black curve is a Curie-Weiss fit from 150 K to 300 K. Inset shows the inverse of magnetic susceptibility. (b) DC magnetic susceptibility from 0.4 K to 4.4 K measured in the same condition as (a). (c) Magnetization (purple, left axis) and its derivative susceptibility (orange, right axis) along [1, 1, 0] direction at 0.7 K.}
	\label{figs3}
\end{figure}

DC magnetic susceptibility was measured in a Quantum Design MPMS [Fig. \ref{figs3}(a)]. No magnetic phase transition can be found from 2 K to 300 K. Curie-Weiss fit from 150 K to 300 K gives a Weiss temperature $\Theta_{\mathrm{cw}}$ = -29.6(1) K [inset of Fig. \ref{figs3}(a)] and effective moment $\mu_{\mathrm{eff}}$ =  3.22(1) $\rm\mu_B/Ni$. Magnetic susceptibility at lower temperature was measured with a He3 option. A sharp peak around 1.1 K can be observed [Fig. \ref{figs3}(b)], which marks an antiferromagnetic phase transition. No additional phase transition can be discerned from 0.4 K to 1 K. This observation can be further confirmed by the temperature dependence of the magnetic Bragg peak (2/3, 2/3, 2/3), which shows no anomaly from 0.1 K to 0.9 K [Fig. \ref{figs1}(b)].

\subsection{Magnetization under pulsed magnetic field}
Magnetization of \ch{K_2Ni_2(SO_4)_3} was measured up to 60 T in the pulsed field facility at Los Alamos National Laboratory by detecting the field-induced voltage across a pick-up coil made of high purity Cu. The sample was carefully aligned along the [1, 1, 0] crystallographic direction, and was put in a 1 mm diameter sample holder. The magnetic field was applied along [1, 1, 0] and magnetization along this direction was measured. For each temperature, two measurements were performed with sample in and out of the pick-up coil. The signal difference between the two was taken as the sample signal which was then calibrated according to the PPMS DC magnetization data. As showed in Fig. \ref{figs3}(c), the saturation field is about 30 T at 0.7 K and the full moment size is $\sim$2.44 $\rm\mu_B/Ni$. Therefore, we can estimate a $g$-factor of 2.44 for \ch{Ni^{2+}} with S = 1. The resultant effective moment is 3.45 $\rm\mu_B$/Ni, which is reasonably consistent with the result obtained from the Curie-Weiss fit.

\section{Absolute intensity normalization and spectral weight}

\begin{figure}[t!]
	\centering{\includegraphics[clip,width=11cm]{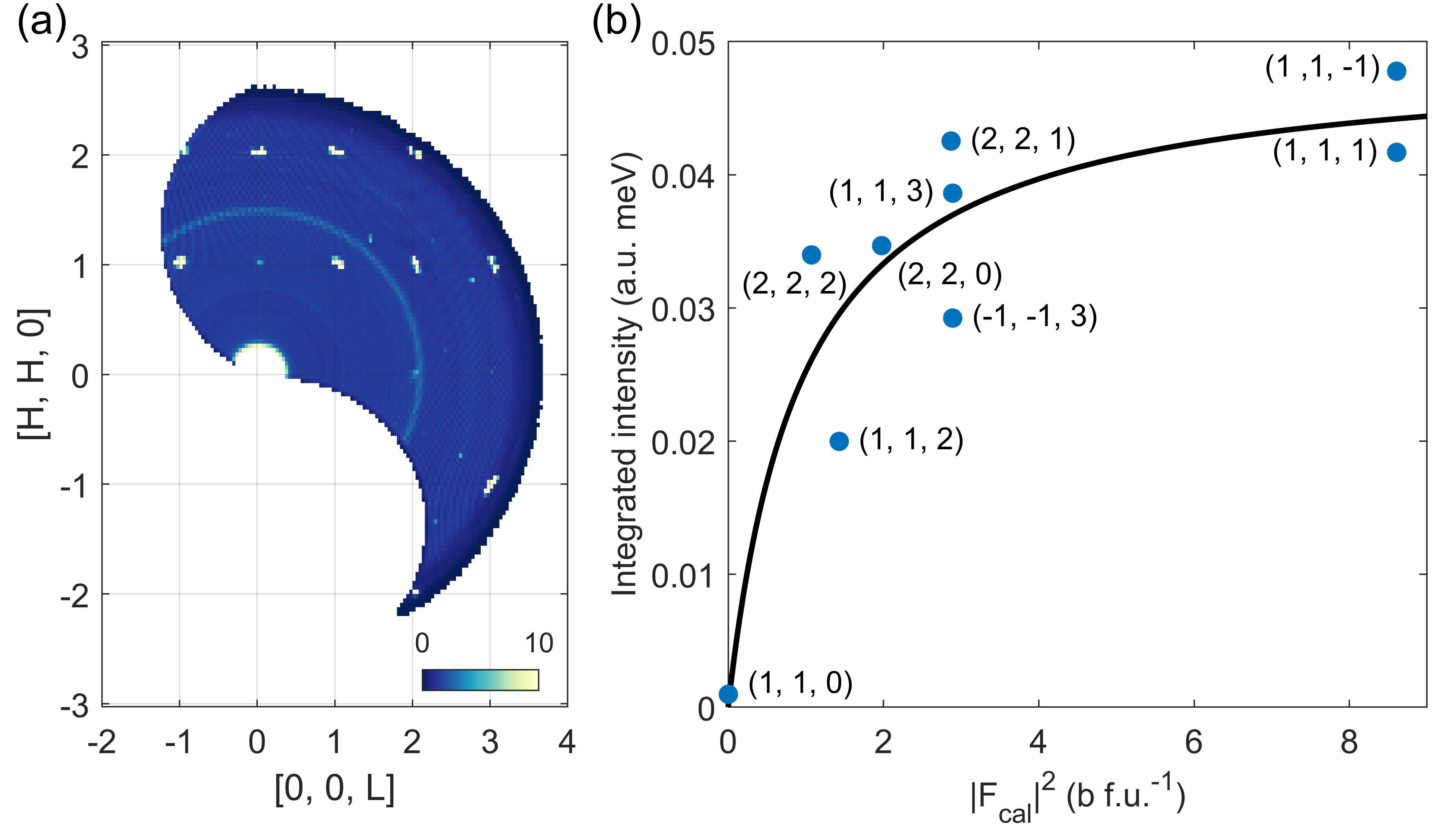}}
	\caption{(a) Elastic scattering pattern of the ($H$, $H$, $L$) plane at 2 K. (b) Calculated neutron scattering structure factor and the corresponding integrated intensity for structural Bragg peaks in (a). The ring that goes through (0, 0, 2) comes from sample environment. Solid line is the fit to the empirical extinction function as described in the text.}
	\label{figs4}
\end{figure}

We converted the measured neutron scattering intensity into absolute cross section with structural Bragg peaks \cite{WuScience2016,MarcusPRL2018,LuoPRB2020}. The elastic scattering pattern of the ($H$, $H$, $L$) plane at 2 K is presented in Fig. \ref{figs4}(a), where 9 Bragg peaks are covered by the measured range without contamination. For a structural Bragg peak, the integrated intensity is proportional to the modulus square of the structure factor $F_{\rm{cal}}(\textbf{Q})$ \cite{Shirane2002}
\begin{equation}
	I(\textbf{Q})=A|F_{\rm{cal}}(\textbf{Q})|^2,
\end{equation}
where
\begin{equation}
	F_{\rm{cal}}(\textbf{Q})=\sum_{j}b_{j}e^{i\textbf{Q}\cdot\textbf{r}_j}e^{-W_j}.
\end{equation}
$F(\textbf{Q})$ can be calculated based on the crystal structure (see Section IB) and coherent neutron scattering length $b_j$ for atoms at site $j$. At low temperature we approximate the Debye–Waller factor ($e^{-W_j}$) to be unity.

In a real scattering process, the measured intensity can be affected by extinction effect, especially for experiments on large crystals. We therefore use the following empirical function to correct the measured intensity \cite{WuScience2016}
\begin{equation}
	I(\textbf{Q})=A\frac{|F_{\rm{cal}}(\textbf{Q})|^2}{1+B|F_{\rm{cal}}(\textbf{Q})|^2}.
\end{equation}
The parameters $A$ and $B$ can be determined by fitting the observed integrated intensity and the modulus square of the calculated structure factor, which are presented in Fig. \ref{figs4}(b). In the limit of $|F_{\rm{cal}}(\textbf{Q})|^2 \rightarrow 0$, (3) reduces to (1), where no extinction effect needs to be considered. According to Fig. \ref{figs4}(b), we obtained $A$ = 0.052 a.u. meV f.u. b$^{-1}$, based on which we can put the measured intensity in absolute unit.

The measured magnteic scattering intensity is then related to the dynamical structure factor $S(\textbf{Q},\omega)$ as
\begin{equation}
	I(\textbf{Q},\omega)/A= \left(\frac{\gamma r_0}{2}\right)^2g^2f^2(Q)e^{-2W}\sum_{\alpha,\beta}{\left(\delta_{\alpha,\beta}-\hat{Q}_{\alpha}\hat{Q}_{\beta}\right)S^{\alpha \beta}(\textbf{Q},\omega)},
\end{equation}
where $\left(\frac{\gamma r_0}{2}\right)^2$ = 0.0726 b and $f(Q)$ is the magnetic form factor of \ch{Ni^{2+}} ion. The dynamic structure factor obeys the total moment sum rule \cite{Shirane2002}
\begin{equation}
	\frac{\sum_{\alpha}\int \mathrm{d}\omega\int_{\mathrm{BZ}}\mathrm{d}\textbf{Q}S^{\alpha \alpha}(\textbf{Q},\omega)}{\int_{\mathrm{BZ}}\mathrm{d}\textbf{Q}}=S(S+1).
\end{equation}
Combing (4) and (5), this sum rule can be approximated to
\begin{equation}
	\frac{10.33 (\mathrm{b}^{-1})\int \mathrm{d}\omega\int_{\mathrm{BZ}}\mathrm{d}\textbf{Q}I(\textbf{Q},\omega)/f^2(Q)/A}{\int_{\mathrm{BZ}}\mathrm{d}\textbf{Q}}=g^2S(S+1),
\end{equation}
% =\left(\frac{\mu_{\mathrm{eff}}}{\mu_{\mathrm{B}}}\right)^2
where the right side is the square of calculated effective magnetic moment per \ch{Ni^{2+}} (in $\mu_{\mathrm{B}}^2$). By performing the integral over the measured Brillouin zones, we get the energy dependence of the intensity as showed in Fig. \ref{figs4}. The spectral weight from 0.1 meV to 2 meV  gives 9.40 $\mu_{\mathrm{B}}^2$ and 9.97 $\mu_{\mathrm{B}}^2$ for 0.1 K and 2 K, respectively, which are close to the square of the effective moment ($\mu_{\mathrm{eff}}^2$ = 10.37 $\mu_{\mathrm{B}}^2$) obtained by magnetic susceptibility (Section IC). This result indicates we recovered most of the spectral weight expected from S = 1 \ch{Ni^{2+}}. Moreover, we can estimate the static moment contributes only $\sim$5.67\% of the total spectral weight, which is much smaller than a ratio of 50\% for conventional magnets. Therefore, most of the spectral weight in \ch{K_2Ni_2(SO_4)_3} is in the dynamical part even at 0.1 K, which is a direct evidence of dominant quantum fluctuations \cite{PlumbNP2019}.

\begin{figure}[h!]
	\centering{\includegraphics[clip,width=9cm]{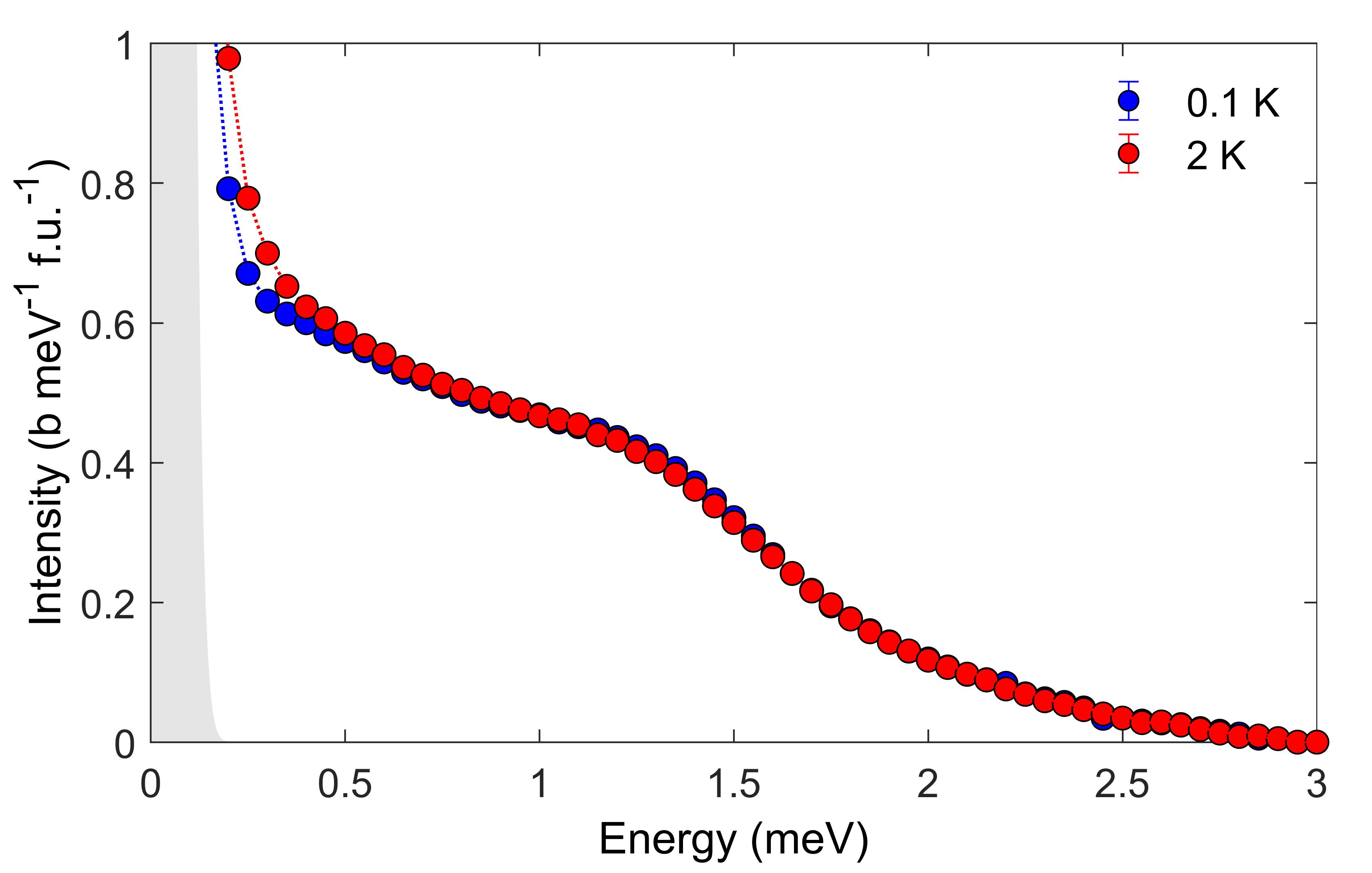}}
	\caption{Energy dependence of the intensity that averaged over Brillouin zones at 0.1 K and 2 K. Shaded area at low energy indicates elastic signal, which is determined according to the energy resolution ($\sim$0.1 meV) and the elastic intensity at 2 K.}
	\label{figs5}
\end{figure}

\section{Additional neutron scattering data}
\begin{figure}[h!]
	\centering{\includegraphics[clip,width=10cm]{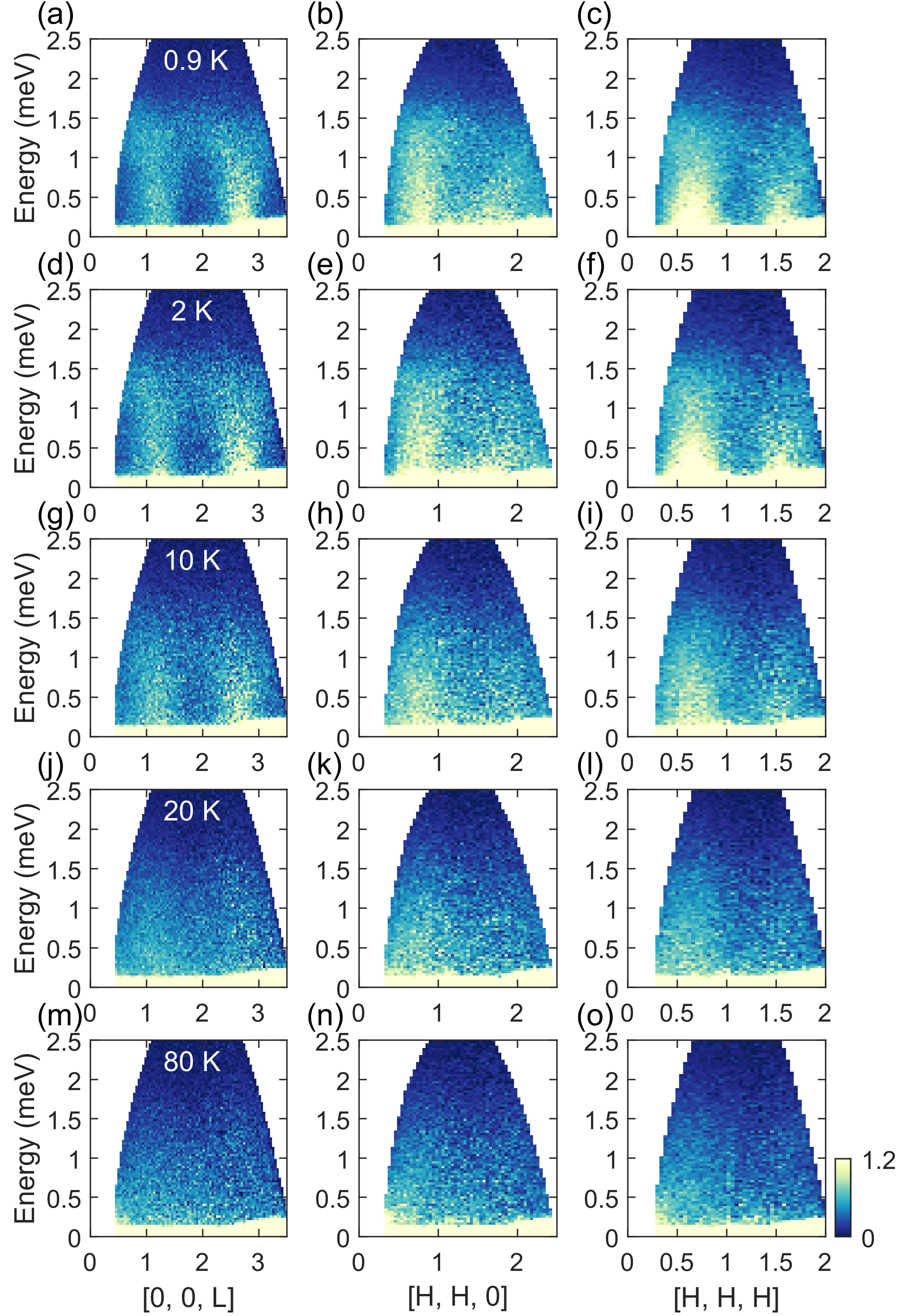}}
	\caption{Energy dependence of the magnetic excitations along high-symmetric directions at 0.9 K [(a)-(c)], 2 K [(d)-(f)], 10 K [(g)-(i)], 20 K [(j)-(l)], and 80 K [(m)-(o)].}
	\label{figs6}
\end{figure}

\begin{figure}[h!]
	\centering{\includegraphics[clip,width=9cm]{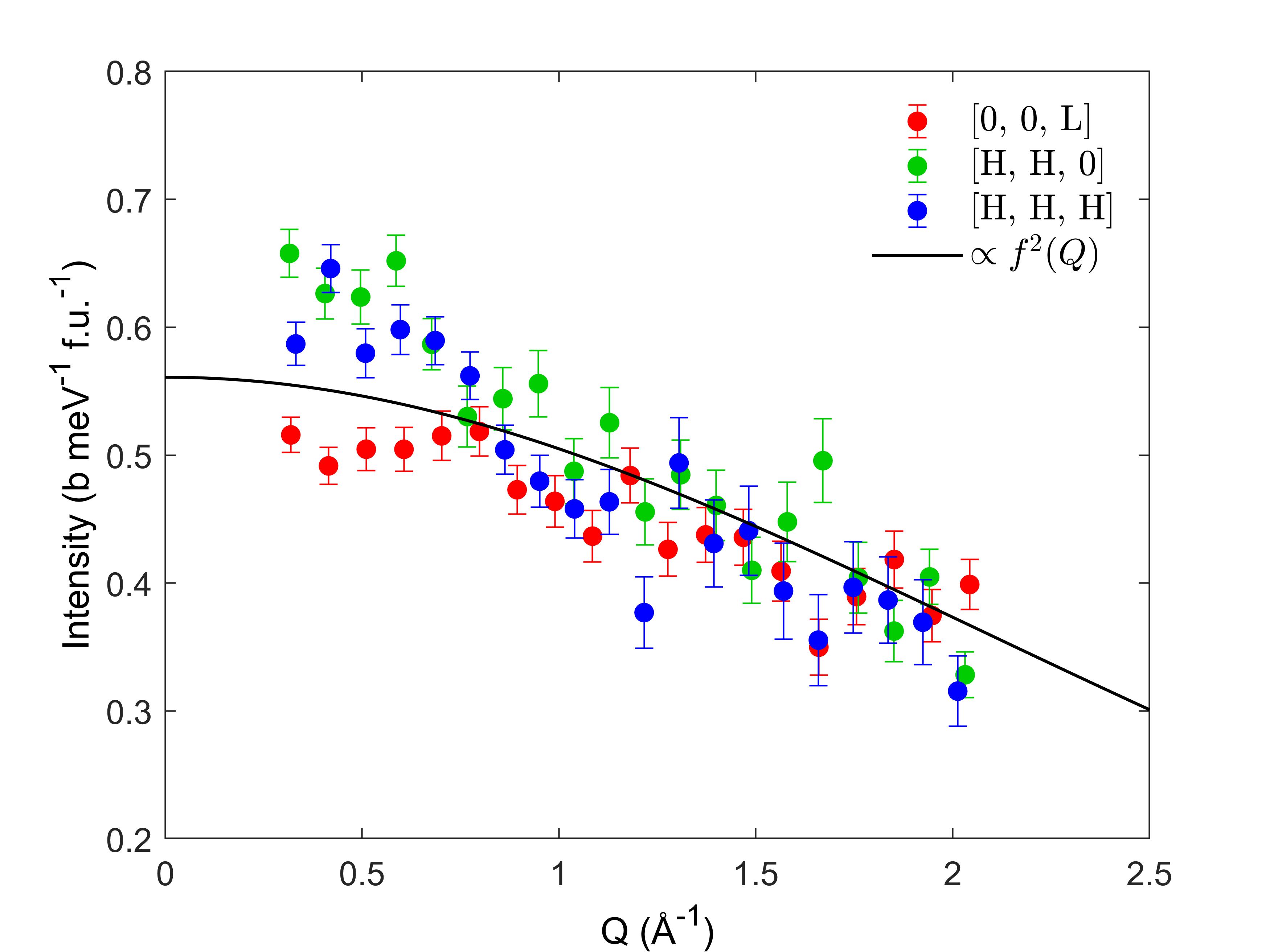}}
	\caption{Q-dependence of the scattering intensity of 0.5 meV at 80 K along three high-symmetric directions. The solid line is a fit to the square of magnetic factor of \ch{Ni^{2+}}.}
	\label{figs7}
\end{figure}

In Fig. \ref{figs6}, we present the magnetic excitation spectra along high-symmetric directions at five temperatures, which supplement the data at 0.1 K showed in the main text. At 80 K, the excitations mainly come from paramagnetic diffuse scattering. The momentum dependence of the intensity at 0.5 meV is showed in Fig. \ref{figs7}, which largely follows the decay of the magnetic form factor.

\section{Self-consistent-gaussian-approximation (SCGA) calculation}

\subsection{Construction of interaction matrix}

The magnetic Hamiltonian can be generally written as \cite{TothJPCM2015}
\begin{equation}
	\begin{aligned}
		H &= \frac{1}{2}\sum_{m,n}\sum_{\mu,\nu}\textbf{S}_{m,\mu}^{\mathrm{T}} \mathrm{J}_{m,\mu;n,\nu} \textbf{S}_{n,\nu}\\
		&= \frac{1}{2}\sum_{m,n}\sum_{\mu,\nu} \begin{pmatrix}
			S_{m,\mu}^x&S_{m,\mu}^y&S_{m,\mu}^z 
		\end{pmatrix}
		\begin{pmatrix}
			J_{m,\mu;n,\nu}^{xx} & J_{m,\mu;n,\nu}^{xy} & J_{m,\mu;n,\nu}^{xz}\\
			J_{m,\mu;n,\nu}^{yx} & J_{m,\mu;n,\nu}^{yy} & J_{m,\mu;n,\nu}^{yz}\\
			J_{m,\mu;n,\nu}^{zx} & J_{m,\mu;n,\nu}^{zy} & J_{m,\mu;n,\nu}^{zz}\\ 
		\end{pmatrix} \begin{pmatrix}
			S_{n,\nu}^x\\S_{n,\nu}^y\\S_{n,\nu}^z
		\end{pmatrix},
	\end{aligned}
\end{equation}
where $\textbf{S}_{m,\mu}$ ($\textbf{S}_{n,\nu}$) represents the 3$\times$1 spin vector of the $\mu$th ($\nu$th) site in the $m$th ($n$th) unit cell ($m, n$ = 1, 2, $\cdots$, N, N is the total number of unit cell; $\mu, \nu$ = 1, 2, $\cdots$, 8), and $\mathrm{J}_{m,\mu;n,\nu}$ is a 3$\times$3 exchange interaction matrix that connects $\textbf{S}_{m,\mu}$ and $\textbf{S}_{n,\nu}$. Here, we consider Heisenberg type exchange interaction
\begin{equation}
	J_{m,\mu;n,\nu}^{\alpha,\beta} = \delta_{\alpha,\beta}J_{m,\mu;n,\nu},
\end{equation}
with $\alpha,\beta = x, y, z$. The Hamiltonian can be further written in the basis of spin vectors in one unit cell (24$\times$1 matrix)
\begin{equation}
	\begin{aligned}
		H &= \frac{1}{2}\sum_{m,n} \begin{pmatrix}
			S_{m,1}^x&S_{m,1}^y&S_{m,1}^z \cdots S_{m,8}^x&S_{m,8}^y&S_{m,8}^z
		\end{pmatrix}
		\mathbb{J}_{m,n}
		\begin{pmatrix}
			S_{n,1}^x\\S_{n,1}^y\\S_{n,1}^z\\ \vdots \\ S_{n,8}^x\\S_{n,8}^y\\S_{n,8}^z\\
		\end{pmatrix},
	\end{aligned}
\end{equation}
where $\mathbb{J}_{m,n}$ is a $24\times24$ interaction matrix
\begin{equation}
	\mathbb{J}_{m,n}=
	\begin{pmatrix}
		J_{m,1;n,1} & 0 & 0 & \cdots & J_{m,1;n,8} & 0 & 0 \\
		0 & J_{m,1;n,1} & 0 & \cdots & 0 & J_{m,1;n,8} & 0 \\
		0 & 0 & J_{m,1;n,1} & \cdots & 0 & 0 & J_{m,1;n,8} \\
		\vdots & \vdots & \vdots & \ddots & \vdots & \vdots & \vdots \\
		J_{m,8;n,1} & 0 & 0 & \cdots & J_{m,8;n,8} & 0 & 0 \\
		0 & J_{m,8;n,1} & 0 & \cdots & 0 & J_{m,8;n,8} & 0 \\
		0 & 0 & J_{m,8;n,1} & \cdots & 0 & 0 & J_{m,8;n,8}
	\end{pmatrix}.
\end{equation}
To proceed, we make Fourier transformation
\begin{equation}
	S_{\mu}^{\alpha}(\textbf{q}) = \frac{1}{N}\sum_{m} e^{-i\textbf{q} \cdot (\textbf{R}_m + \textbf{r}_{\mu})} S_{m,\mu}^{\alpha},
\end{equation}
where $\textbf{R}_m$ is the position of $m$th unit cell, $\textbf{r}_{\mu}$ is the relative position of the spin component within this unit cell, and the summation on $m$ is taken over the number of unit cells (1, 2, $\cdots$, N). The Hamiltonian can be further written in momentum space as
\begin{equation}
	\begin{aligned}
		H &= \frac{1}{2}\sum_{\textbf{q}} \begin{pmatrix}
			S_{1}^x(\textbf{q})&S_{1}^y(\textbf{q})&S_{1}^z(\textbf{q}) \cdots S_{8}^x(\textbf{q})&S_{8}^y(\textbf{q})&S_{8}^z(\textbf{q})
		\end{pmatrix}
		\mathbb{J}(\textbf{q})
		\begin{pmatrix}
			S_{1}^x(-\textbf{q})\\S_{1}^y(-\textbf{q})\\S_{1}^z(-\textbf{q})\\ \vdots \\ S_{8}^x(-\textbf{q})\\S_{8}^y(-\textbf{q})\\S_{8}^z(-\textbf{q})
		\end{pmatrix}.
	\end{aligned}
\end{equation}
The interaction matrix is now
\begin{equation}
	\mathbb{J}(\textbf{q})=
	\begin{pmatrix}
		J_{1;1}(\textbf{q}) & 0 & 0 & \cdots & J_{1;8}(\textbf{q}) & 0 & 0 \\
		0 & J_{1;1}(\textbf{q}) & 0 & \cdots & 0 & J_{1;8}(\textbf{q}) & 0 \\
		0 & 0 & J_{1;1}(\textbf{q}) & \cdots & 0 & 0 & J_{1;8}(\textbf{q}) \\
		\vdots & \vdots & \vdots & \ddots & \vdots & \vdots & \vdots \\
		J_{8;1}(\textbf{q}) & 0 & 0 & \cdots & J_{8;8}(\textbf{q}) & 0 & 0 \\
		0 & J_{8;1}(\textbf{q}) & 0 & \cdots & 0 & J_{8;8}(\textbf{q}) & 0 \\
		0 & 0 & J_{8;1}(\textbf{q}) & \cdots & 0 & 0 & J_{8;8}(\textbf{q})
	\end{pmatrix},
\end{equation}
with the matrix element as
\begin{equation}
	J_{\mu;\nu}(\textbf{q}) = e^{-i\textbf{q} \cdot (\textbf{r}_{\mu} - \textbf{r}_{\nu})} \sum_{\Delta \textbf{R}} e^{-i\textbf{q} \cdot \Delta \textbf{R}} J_{\mu;\nu}(\Delta \textbf{R}),
\end{equation}
where $J_{\mu;\nu}(\Delta \textbf{R})$ is obtained by reorganizing the exchange interaction $J_{m,\mu;n,\nu}$ between spins at $\textbf{R}_m + \textbf{r}_{\mu}$ and $\textbf{R}_n + \textbf{r}_{\nu}$. We note that the exchange interaction between two spins only depends on their relative disctance $\Delta \textbf{R} + \textbf{r}_{\mu} - \textbf{r}_{\nu}$ with $\Delta \textbf{R} = \textbf{R}_m - \textbf{R}_m$. $\mathbb{J}(\textbf{q})$ has the following eigen equation
\begin{equation}
	\mathbb{J}(\textbf{q})\,\textbf{f}_s(\textbf{q})=\omega_s(\textbf{q})\, \textbf{f}_s(\textbf{q}),
\end{equation}
where $\textbf{f}_s(\textbf{q})$ is the eigen vector that corresponds to the $s$th eigen value $\omega_s(\textbf{q})$, with $s$ = 1, 2, $\cdots$, 24.

\subsection{Calculation and fitting of the neutron scattering intensity}

Within self-consistent-gaussian-approximation (SCGA) method, the spin-spin correlation can be expressed as \cite{ConlonPRB2010,BentonJPSJ2015,PlumbNP2019,BaiPRL2019,PaddisonPRL2020,GuPRB2022,GaoPRL2022,GaoPRB2022}
\begin{equation}
	\langle S_{\mu}^{\alpha}(-\textbf{q})S_{\nu}^{\beta}(\textbf{q})\rangle = \sum_{s} \frac{f_s^{\mu,\alpha}(\textbf{q})^{*}\,f_s^{\nu,\beta}(\textbf{q})}{\lambda+\beta\omega_{s}(\textbf{q})}.
\end{equation}
$\beta$ in the right side is 1/$k_\mathrm{B}T$, with $k_\mathrm{B}$ being Boltzmann constant and $T$ being temperature. $f_s^{\mu,\alpha}(\textbf{q})$ [$f_s^{\nu,\beta}(\textbf{q})$] is the element of $\textbf{f}_s(\textbf{q})$ that corresponds to the spin component $\alpha$ ($\beta$) at site $\mu$ ($\nu$). $\lambda$ is a factor that needs to be determined self-consistently to satisfy the following spin length constrain \cite{ConlonPRB2010,BentonJPSJ2015,PlumbNP2019,BaiPRL2019,PaddisonPRL2020,GuPRB2022,GaoPRL2022,GaoPRB2022}
\begin{equation}
	\sum_{s}\sum_{\textbf{q}}\frac{1}{\lambda+\beta\omega_{s}(\textbf{q})} = 8N,
\end{equation}
where the summation over $\textbf{q}$ is evenly taken on $N$ positions at the Brillouin zone. Then, the energy-integrated intensity measured in unpolarized neutron scattering is
\begin{equation}
	I(\textbf{q})= Cf^2(q)\sum_{\mu,\nu}\sum_{\alpha,\beta} \left(\delta_{\alpha,\beta}-\hat{q}_{\alpha}\hat{q}_{\beta}\right) \langle S_{\mu}^{\alpha}(-\textbf{q})S_{\nu}^{\beta}(\textbf{q})\rangle,
\end{equation}
where $C$ is a constant.

\begin{figure}[b!]
	\centering{\includegraphics[clip,width=16cm]{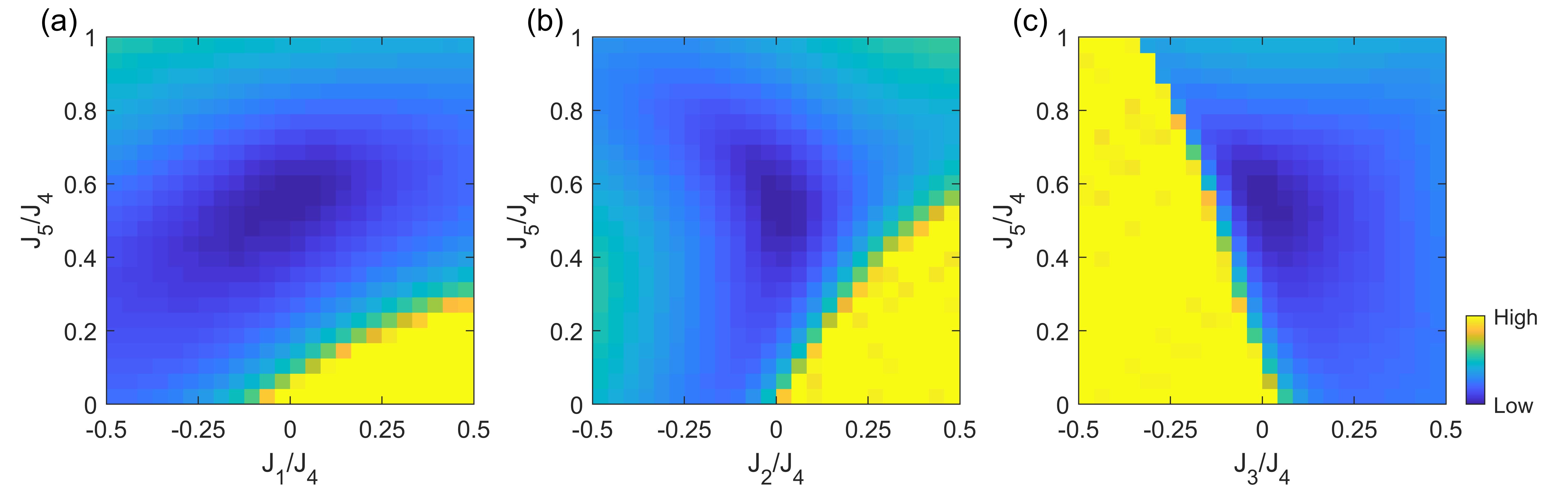}}
	\caption{Maps of log($\chi^2$) for $J_1$, $J_2$, $J_3$, and $J_5$ obtained in the SCGA fitting. The dominant exchange interaction $J_4$ is fixed at the optimized value and is used as the energy unit.}
	\label{figs8}
\end{figure}

Based on SCGA, we fitted the experiment data with the five nearest neighbor exchange interactions ($J_1$, $J_1$, $\cdots$, $J_5$) (see Table \ref{tb3}). The overall magnitude of $J$s cannot be uniquely constrained by the SCGA alone because it only adjusts an overall intensity scale factor that is correlated with temperature \cite{PlumbNP2019,GuPRB2022}. We therefore used the exchange interactions obtained by DFT calculation \cite{ZivkovicPRL2021} as the starting point, which give a Weiss temperature ($|\Theta_{\rm{CW}}|$ = 25.2 K) that is close to experiment. The optimal parameters reported in the main text were determined by least square method, with the goodness-of-fit being defined as
\begin{equation}
	\chi^2=\sum_{\textbf{q}}|I_{\rm{obs}}(\textbf{q})-I_{\rm{cal}}(\textbf{q})|^2,
\end{equation}
where $I_{\rm{obs}}(\textbf{q})$ and $I_{\rm{cal}}(\textbf{q})$ are observed and calculated intensities, respectively, and the summation is taken over the measured momentum positions in the ($H$, $H$, $L$) plane. Fig. \ref{figs8} presents the maps of log($\chi^2$) obtained in the SCGA fitting.

\subsection{About single-ion anisotropy}

\begin{figure}[b!]
	\centering{\includegraphics[clip,width=15cm]{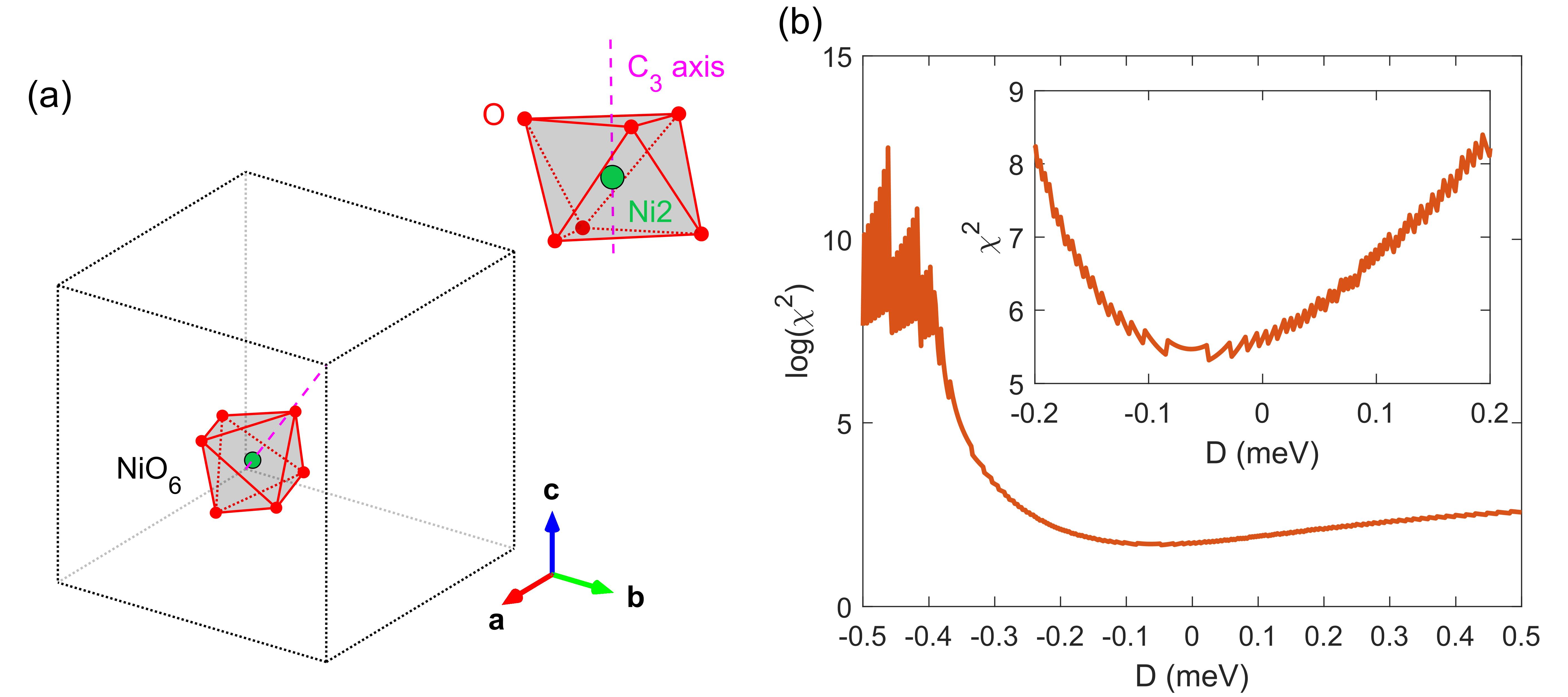}}
	\caption{(a) One representative \ch{NiO_6} octahedron (for Ni2) in the cubic unit cell. The dashed magenta line indicates the [1, 1, 1] direction, which is the C$_3$ axis of this octahedron. The upper right part highlights the \ch{NiO_6} octahedron with the C$_3$ axis being put vertical. (b) Evolution of log($\chi^2$) with respect to the single-ion anisotropy of \ch{Ni^{2+}}. The inset shows the details of $\chi^2$ when $D$ is close to zero.}
	\label{figs9}
\end{figure}

For spin size larger than 1/2, single-ion anisotropy is generally allowed, which can be written as
\begin{equation}
	H_{\rm{aniso}}=\sum_{i}D_i(\textbf{S}_i\cdot\textbf{n}_i)^2,
\end{equation}
where $D_i$ is the magnitude of single-ion anisotropy for the $i$-th Ni site (with $i = 1 \cdots 8$) and $\textbf{n}_i$ is the unit vector of corresponding single-ion axis. $D_i < 0$ and $D_i > 0$ represent the cases of easy-axis and easy-plane, respectively. In \ch{K_2Ni_2(SO_4)_3}, each \ch{NiO_6} octahedron has a C$_3$ axis along the body diagonal of the cubic unit cell [Fig. \ref{figs9} (a)]. Allowed C$_3$ axes in the global frame are
\begin{equation}
	\left\{
	\begin{aligned}
		\textbf{n}_1 = \textbf{n}_5 = &\frac{1}{\sqrt{3}}(1, 1, 1),&\\
		\textbf{n}_2 = \textbf{n}_6 = &\frac{1}{\sqrt{3}}(-1, -1, 1),&\\
		\textbf{n}_3 = \textbf{n}_7 = &\frac{1}{\sqrt{3}}(-1, 1, -1),&\\
		\textbf{n}_4 = \textbf{n}_8 = &\frac{1}{\sqrt{3}}(1, -1, -1).&\\
	\end{aligned}
	\right.
\end{equation}
After Fourier transformation, the anisotropic term can be explicitly written as
\begin{equation}
	\begin{aligned}
		H_{\rm{aniso}}&= \frac{1}{2}\sum_{\textbf{q}} \begin{pmatrix}
			S_{1}^x(\textbf{q})&S_{1}^y(\textbf{q})&S_{1}^z(\textbf{q}) \cdots S_{8}^x(\textbf{q})&S_{8}^y(\textbf{q})&S_{8}^z(\textbf{q})
		\end{pmatrix}
		\mathbb{D}(\textbf{q})
		\begin{pmatrix}
			S_{1}^x(-\textbf{q})\\S_{1}^y(-\textbf{q})\\S_{1}^z(-\textbf{q})\\ \vdots \\ S_{8}^x(-\textbf{q})\\S_{8}^y(-\textbf{q})\\S_{8}^z(-\textbf{q})
		\end{pmatrix},
	\end{aligned}
\end{equation}
where $\mathbb{D}(\textbf{q})$ is the single-ion anisotropy matrix
\begin{equation}
	\mathbb{D}(\textbf{q})=\frac{2}{3}
	\begin{pmatrix}
		D_1 & D_1 & D_1 & \cdots & 0 & 0 & 0 \\
		D_1 & D_1 & D_1 & \cdots & 0 & 0 & 0 \\
		D_1 & D_1 & D_1 & \cdots & 0 & 0 & 0 \\
		\vdots & \vdots & \vdots & \ddots & \vdots & \vdots & \vdots \\
		0 & 0 & 0 & \cdots & D_8 & -D_8 & -D_8 \\
		0 & 0 & 0 & \cdots & -D_8 & D_8 & D_8 \\
		0 & 0 & 0 & \cdots & -D_8 & D_8 & D_8
	\end{pmatrix}.
\end{equation}
$\mathbb{D}(\textbf{q})$ can be written in a block-diagonal form
\begin{equation}
	\mathbb{D}(\textbf{q})=2D
	\begin{pmatrix}
		\textbf{n}_1^{\rm{T}}\cdot\textbf{n}_1&\textbf{0}&\textbf{0}&\textbf{0}&\textbf{0}&\textbf{0}&\textbf{0}&\textbf{0}\\
		\textbf{0}&\textbf{n}_2^{\rm{T}}\cdot\textbf{n}_2&\textbf{0}&\textbf{0}&\textbf{0}&\textbf{0}&\textbf{0}&\textbf{0}\\
		\textbf{0}&\textbf{0}&\textbf{n}_3^{\rm{T}}\cdot\textbf{n}_3&\textbf{0}&\textbf{0}&\textbf{0}&\textbf{0}&\textbf{0}\\
		\textbf{0}&\textbf{0}&\textbf{0}&\textbf{n}_4^{\rm{T}}\cdot\textbf{n}_4&\textbf{0}&\textbf{0}&\textbf{0}&\textbf{0}\\
		\textbf{0}&\textbf{0}&\textbf{0}&\textbf{0}&\textbf{n}_5^{\rm{T}}\cdot\textbf{n}_5&\textbf{0}&\textbf{0}&\textbf{0}\\
		\textbf{0}&\textbf{0}&\textbf{0}&\textbf{0}&\textbf{0}&\textbf{n}_6^{\rm{T}}\cdot\textbf{n}_6&\textbf{0}&\textbf{0}\\
		\textbf{0}&\textbf{0}&\textbf{0}&\textbf{0}&\textbf{0}&\textbf{0}&\textbf{n}_7^{\rm{T}}\cdot\textbf{n}_7&\textbf{0}\\
		\textbf{0}&\textbf{0}&\textbf{0}&\textbf{0}&\textbf{0}&\textbf{0}&\textbf{0}&\textbf{n}_8^{\rm{T}}\cdot\textbf{n}_8\\
	\end{pmatrix},
\end{equation}
where we have assumed same magnitude of single-ion anisotropic term ($D$) for all eight Ni sites and $\textbf{0}$ is a $3\times3$ zero matrix. The SCGA calculation procedure can be applied with the eigen values and eigen states of $\mathbb{J}(\textbf{q})+\mathbb{D}(\textbf{q})$.

With the inclusion of a single-ion anisotropic term, the fitting cannot be significantly improved. In more specific, if we further include $\mathbb{D}(\textbf{q})$ in the calculation, the resultant $\chi^2$ has a minimum close to $D$ = 0, as showed in Fig. \ref{figs9} (b). The exchange interactions of $\mathbb{J}(\textbf{q})$ are fixed at those parameters reported in the main text. In Fig. \ref{figs9} (b), the actual minimum of $\chi^2$ arrives when $D \approx -0.05 $ meV, which is only about 10\% of the dominant exchange interaction $J_4$. Therefore, based on our experiment data, we consider the single-ion anisotropy of \ch{Ni^{2+}} only has a minor effect. Similar conclusion was made in \ch{NaCaNi_2F_7} with a pyrochlore lattice \cite{PlumbNP2019}.

\subsection{$J_5$-only and $J_4$-only cases}

\begin{figure}[h!]
	\centering{\includegraphics[clip,width=14cm]{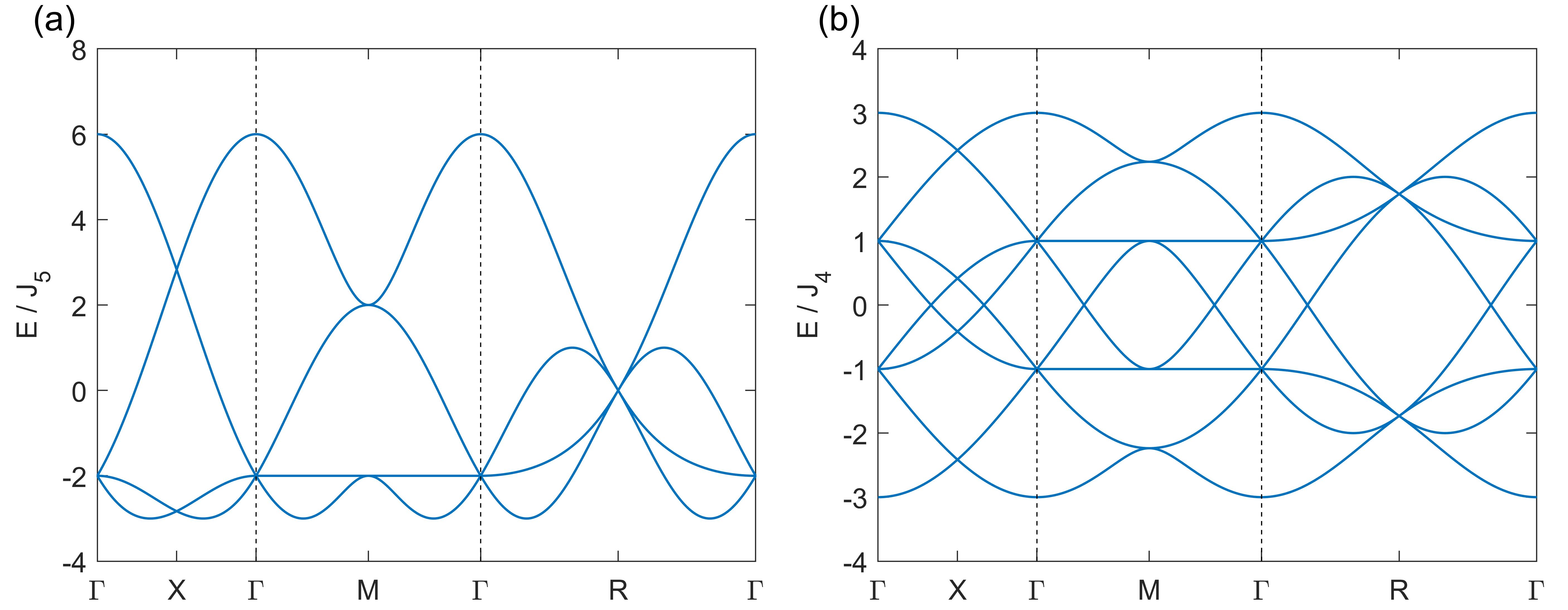}}
	\caption{Energy bands along high-symmetric directions of the Brillouin zone [see Fig. 5(e) of the main text] for cases of $J_5$-only (a) and $J_4$-only (b), where the energy is in the unit of $J_5$ and $J_4$, respectively.}
	\label{figs10}
\end{figure}

To see the effect of coexistence of inter- and intra-trillium-lattice interactions, Fig. \ref{figs10} show the eigen values of the interaction matrix for $J_5$-only and $J_4$-only cases. We can see that separated ``flat bands'' are absent. The $J_5$-only case in Fig. \ref{figs10} (a) corresponds to a single trillium lattice, which is the same as the result reported in \cite{HopkinsonPRB2006}.

\begin{table}[h]
	\caption{
		Crystal data and structure refinement for \ch{K_2Ni_2(SO_4)_3}.
	}
	\begin{ruledtabular}
		\begin{tabular}{cc}
			Empirical formula&\ch{K_2Ni_2O_12S_3}\\
			Formula weight& 483.80\\
			Temperature&301(2) K\\
			Wavelength&0.71073 \AA\\
			Crystal system&Cubic\\
			Space group&$P$2$_1$3\\
			Unit cell dimensions&a = b = c = 9.83610(12) \AA\\
			&$\alpha$ = $\beta$ = $\gamma$ = 90$^{\circ}$\\
			Volume&951.63(3)\AA$^3$\\
			Z&4\\
			Density (calculated)&3.377 Mg/m$^3$\\
			Absorption coefficient&5.559 mm$^{-1}$\\
			F(000)&952\\
			Crystal size&0.143 $\times$ 0.142 $\times$ 0.074 mm$^3$\\
			Theta range for data collection&2.929 to 35.472$^{\circ}$\\
			Index ranges&-15 $\le$ h $\le$ 13, -15 $\le$ k $\le$ 12, -14 $\le$ l $\le$ 11\\
			Reflections collected&13616\\
			Independent reflections&1374 [R(int) = 0.0404]\\
			Completeness to theta = 25.242$^{\circ}$&100.0 \%\\
			Absorption correction&None\\
			Refinement method&Full-matrix least-squares on F$^2$\\
			Data / restraints / parameters&1374 / 0 / 59\\
			Goodness-of-fit on F$^2$&1.068\\
			Final R indices [I $>$ 2sigma(I)]&R1 = 0.0498, wR2 = 0.0982\\
			R indices (all data)&R1 = 0.0543, wR2 = 0.0999\\
			Absolute structure parameter&0.954(7)\\
			Extinction coefficient&0.0137(13)\\
			Largest diff. peak and hole&0.642 and -0.492 e.\AA$^{-3}$\\
		\end{tabular}	
	\end{ruledtabular}
	\label{tb1}
\end{table}

\begin{table}[h]
	\caption{Atomic coordinates and equivalent isotropic displacement parameters of \ch{K_2Ni_2(SO_4)_3} at 301(2) K. [U$\rm_{eq}$ is defined as one-third of the trace of the orthogonalized U$\rm_{ij}$ tensor (\AA$^2$)].
	}
	\begin{ruledtabular}
		\begin{tabular}{ccccccc}
			\textbf{Atom}&\textbf{Wyckoff.}&\textbf{Occ.}&\textbf{x}&\textbf{y}&\textbf{z}&\textbf{U$\rm_{eq}$}\\
			\midrule
			Ni1&4$a$&1&0.3452(1)&0.3452(1)&0.3452(1)&0.0085(3)\\
			Ni2&4$a$&1&0.0854(1)&0.0854(1)&0.0854(1)&0.0083(3)\\
			S3&12$b$&1&0.0330(2)&0.2690(2)&0.3737(2)&0.0068(3)\\
			K4&4$a$&1&0.5648(2)&0.5648(2)&0.5648(2)&0.0191(5)\\
			K5&4$a$&1&0.7987(2)&0.7987(2)&0.7987(2)&0.0200(5)\\
			O6&12$b$&1&0.0027(6)&0.4113(5)&0.3449(7)&0.0202(11)\\
			O7&12$b$&1&0.1746(5)&0.2531(6)&0.4199(6)&0.0163(10)\\
			O8&12$b$&1&0.0233(6)&0.0583(6)&0.7118(6)&0.0209(11)\\
			O9&12$b$&1&0.0094(6)&0.1918(7)&0.2482(6)&0.0239(13)\\
		\end{tabular}	
	\end{ruledtabular}
	\label{tb2}
\end{table}

\begin{table}[h]
	\caption{Bond distance and angle information for \ch{Ni^{2+}} ions.}
	\begin{ruledtabular}
		\begin{tabular}{ccccc}
			Bond&Distance (\AA)&Connection&\makecell{Number of atoms \\for smallest loop}&Ni-S-Ni bond angle ($^{\circ}$)\\
			\midrule
			$J_1$&4.4272&Ni1 - Ni2&2&84.55\\
			$J_2$&4.8995&Ni1 - Ni2&10&97.35\\
			$J_3$&6.0823&Ni1 - Ni1&3&130.90\\
			$J_4$&6.1198&Ni1 - Ni2&10&124.51\\
			$J_5$&6.1257&Ni2 - Ni2&3&130.28\\
			$J_6$&8.1316&Ni1 - Ni2&6&-\footnote{Starting from $J_6$, simple exchange path through one \ch{SO_4^{2-}} group becomes unavailable.}\\
			$J_7$&8.4024&Ni2 - Ni2&3&-\\
			$J_8$&8.5939&Ni1 - Ni1&3&-\\
			$J_9$&8.6202&Ni1 - Ni2&10&-\\
			$J_{10}$&9.3678&Ni1 - Ni2&10&-\\
		\end{tabular}	
	\end{ruledtabular}
	\label{tb3}
\end{table}

\end{document}